# Liquidity-adjusted Return and Volatility, and Autoregressive Models

Qi Deng[1,2]*
Zhong-guo Zhou[3]

## Abstract

We construct liquidity-adjusted return and volatility using purposely designed liquidity metrics (liquidity jump and liquidity diffusion) that incorporate additional liquidity information. Based on these measures, we introduce a liquidity-adjusted ARMA-GARCH framework to address the limitations of traditional ARMA-GARCH models, which are not effectively in modeling illiquid assets with high liquidity variability, such as cryptocurrencies. We demonstrate that the liquidity-adjusted model improves model fit for cryptocurrencies, with greater volatility sensitivity to past shocks and reduced volatility persistence of erratic past volatility. Our model is validated by the empirical evidence that the liquidity-adjusted mean-variance (LAMV) portfolios outperform the traditional mean-variance (TMV) portfolios.

Funding Source: The work was supported by Hubei University of Automotive Technology [grant number BK202209] and Hubei Provincial Bureau of Science and Technology [grant number 2023EHA018].

1. College of Artificial Intelligence, Hubei University of Automotive Technology, Shiyan, China
2. Cofintelligence Financial Technology Ltd., Hong Kong and Shanghai, China
3. Department of Finance, Financial Planning, and Insurance, Nazarian College of Business and Economics, California State University Northridge, CA, USA
*. Corresponding author: dq@huat.edu.cn; qi.deng@cofintelligence.com

# Liquidity-adjusted Return and Volatility, and Autoregressive Models

## 1. Introduction

The Autoregressive Moving Average Generalized Autoregressive Conditional Heteroskedasticity (ARMA-GARCH) framework is a widely used to model financial econometrics for modeling both the conditional mean and volatility of asset returns. However, traditional ARMA-GARCH models have a known limitation, as in general they do not explicitly account for the impact of liquidity fluctuation, which may result in systematically biased parameter estimates and suboptimal risk management strategies. This issue is particularly notable for illiquid assets that exhibit higher liquidity variability (e.g., cryptocurrencies).

To address this limitation, extensions of the ARMA-GARCH framework have been developed by incorporating discrete jump components, resulting in ARMA-GARCH-Jump models (e.g., Chan & Maheu, 2002; Duan, Ritchken & Sun, 2006). However, these models tend to treat all jumps indifferently and do not distinguish between jumps caused by sudden liquidity shocks and those caused by other factors (e.g., macroeconomic news or market sentiment, etc.), as they do not explicitly incorporate key liquidity measures (e.g., trading volume, illiquidity ratio, bid-ask spread, etc.). Furthermore, they often do not capture the interaction between liquidity and volatility, such as the feedback loop between declining liquidity and increasing volatility. These limitations are particularly pronounced for cryptocurrencies. Unlike traditional assets such as U.S. stocks, cryptocurrencies are characterized by extreme volatility and frequent liquidity dislocations. They often experience sudden price movements due to large trades, lack of regulations, or technological developments. These movements are exacerbated by the relatively low liquidity of many cryptocurrencies, which can lead to significant price impacts and increased volatility.

Traditional ARMA-GARCH (and subsequent ARMA-GARCH-Jump) models are not effective for cryptocurrencies for several reasons. First, these models are developed under an implicit assumption that markets are efficient and liquid, which is often not the case for a majority of cryptocurrencies. Cryptocurrency markets are highly fragmented, with trading occurring across multiple exchanges that may have different liquidity conditions and price levels. This fragmentation makes it difficult to apply traditional models, which are typically calibrated for centralized, liquid markets. Second, liquidity risk in cryptocurrencies is driven by factors such as wide swings of trading volume, unstable bid-ask spreads, and sometimes the absence of market makers, which can lead to significant sudden and discontinuous price impacts and increased volatility. As traditional models do not explicitly incorporate liquidity risk, they cannot easily absorb such discontinuities. Third, cryptocurrency markets are largely unregulated, where ill-intentioned practices such as wash trading (e.g., Cong et al., 2023) are not actively monitored and curbed. These manipulative trading practices may distort the true liquidity conditions, compromising the volatility estimate accuracy of traditional models that are designed for regulated markets with mostly transparent trading. Furthermore, traditional models do not account for the high-frequency and continuous nature of cryptocurrency trading. Cryptocurrencies are traded 24/7 and prices can change at high-frequency. Traditional models, typically estimated using daily or lower-frequency data, are ill-equipped to capture high-frequency and intraday dynamics.

In this paper, we address the abovementioned deficiencies by proposing a liquidity-adjusted ARMA-GARCH framework. Specifically, we use the minute-level trading data of selected assets (US stocks and cryptocurrencies) to establish a scaled minute-level illiquidity ratio and derive the minute-level liquidity-adjusted return and volatility, based on which, we calculate the daily-level liquidity-adjusted return and (intraday) volatility, and then apply the ARAM-GARCH to model the liquidity-adjusted return. With the liquidity-adjusted ARMA-GARCH framework, we provide a more precise characterization of liquidity risk, and a more robust econometric representation of the interaction between liquidity risk and volatility dynamics for cryptocurrencies, as well as other asset classes with high liquidity volatility.[1]

To validate the proposed liquidity-adjusted ARMA-GARCH model, we apply it to two distinct asset classes: US stocks as a benchmark representing traditional assets with high liquidity and low liquidity variability, and cryptocurrencies that represent assets with low liquidity and high liquidity variability. We find that while the liquidity-adjusted ARMA-GARCH model does not improve upon the traditional ARMA-GARCH model in volatility estimation for stocks, it indeed yields significant performance enhancement for cryptocurrencies. Specifically, we find that the liquidity-adjust model amplifies the impact of recent, past shocks on volatility, while reduces the estimated persistence of volatility, suggesting that liquidity-adjusted returns better account for transitory liquidity shocks in cryptocurrency trading. Our findings align with the fragmented nature of cryptocurrency markets, where liquidity shocks are more frequent and severe, and that liquidity risk exerts a greater influence on short-term volatility than on long-term volatility persistence. The results underscore the necessity of modeling cryptocurrencies through liquidity-adjusted measures rather than their regular counterparts, as the former incorporates the impact of liquidity risk, which is often underrepresented by the latter.

---

[1] We apply the standard ARMA-GARCH to model the conditional liquidity-adjusted return and volatility, which is essentially a new ARMA-GARCH model adjusted with liquidity measures that we name the "liquidity-adjusted ARMA-GARCH" model. We also use the standard ARMA-GARCH to model the conditional regular return and volatility, which we call the "traditional ARMA-GARCH" model. We use these terms throughout this paper.

In order to provide empirical support to substantiate the effectiveness of the liquidity-adjusted ARMA-GARCH model for cryptocurrencies, for each asset (stock or crypto), we apply the returns predicted by the liquidity-adjusted ARMA-GARCH model as inputs to a mean-variance (MV) portfolio for performance optimization (LAMV), with each portfolio consisting two assets, either a stock or a cryptocurrency, and a risk-free asset. For comparison, we duplicate the procedure with the returns predicted by the traditional ARMA-GARCH model being fed to a MV portfolio (TMV).[2] For US stocks, we do not observe a portfolio performance improvement of the LAMV over the TMV, suggesting that high-liquidity assets inherently reflect liquidity risk within their price dynamics, and therefore that liquidity adjustment is not particularly beneficial for stocks. Conversely, we observe a clear advantage for the LAMV over the TMV in portfolio performance for cryptocurrencies, except for Bitcoin (BTC), the largest cryptocurrency in market cap and liquidity. The results suggest that liquidity adjustment is particularly beneficial for cryptocurrencies, especially those with lower baseline liquidity levels. The empirical evidence supports our proposition that liquidity adjustment restores the effectiveness of ARMA-GARCH in modeling cryptocurrencies, and in general, assets with low liquidity and high liquidity variability.

Our study makes several important contributions to the financial econometrics literature. First, we introduce liquidity jump and liquidity diffusion as novel liquidity measures that capture both the magnitude and intraday volatility of liquidity shocks, refining existing liquidity risk modeling approaches. Second, it extends the traditional ARMA-GARCH framework by explicitly incorporating these liquidity measures, demonstrating that the liquidity-adjusted ARMA-GARCH model significantly improves volatility estimation for cryptocurrencies. Third, the comparative analysis of U.S. stocks and cryptocurrencies highlights the distinct responses of liquidity-adjusted ARMA-GARCH model to asset returns with different levels of embedded liquidity information, underscoring the importance of tailoring risk management strategies to the unique characteristics of each asset class. Fourth, our findings have practical implications for investment, emphasizing the need to account for liquidity risk in portfolio optimization and risk management, particularly for portfolios exposed to illiquid and volatile assets.

---

[2] We use the standard MV construct to optimize portfolios with predicted returns from the liquidity-adjusted ARMA-GARCH model and name these portfolios the "liquidity-adjusted Mean Variance (LAMV)" portfolios. We also use the standard MV to optimize portfolios with predicted returns from the traditional ARMA-GARCH model and name these portfolios the "traditional Mean Variance (TMV)" portfolios. We use these terms throughout this paper.

The rest of the paper proceeds as follows. Section 2 reviews existing literature on liquidity and volatility of liquidity, autoregressive models, assets with low liquidity and high liquidity variability, and identifies research gaps. Section 3 introduces liquidity-adjusted return and volatility, and the liquidity measures (liquidity jump and liquidity diffusion). Section 4 provides the descriptive statistics of the data and discussions on the ensemble distribution of return, volatility, and liquidity. Section 5 presents the liquidity-adjusted ARMA-GARCH model, and demonstrate its improved performance for cryptocurrencies, which is supported by Section 6 that provides empirical evidence to that the LAMV portfolios offer clear performance advantage over the TMV portfolios for cryptocurrencies. Section 7 concludes the paper.

## 2. Literature Review

**2.1 Liquidity Risk and Liquidity Premium Models**

Amihud and Mendelson (1986) define asset illiquidity as the cost of immediate execution (trading friction) and use bid-ask spread as a general measure of illiquidity. Fong, Holden and Trzcinka (2017) identify a range of liquidity proxies and argue that the daily version of Closing Percent Quoted Spread (Chung and Zhang, 2014) is the best daily percent-cost proxy, and that the daily version of illiquidity ratio (Amihud, 2002) the best daily cost-per-dollar-volume proxy.

Acharya and Pedersen (2005) present a simple equilibrium model, the liquidity-adjusted capital asset pricing model, which suggests that liquidity risk is a priced factor in asset pricing. Amihud, Mendelson and Pedersen (2005) classify sources of illiquidity, including exogenous transaction cost, exogenous trading horizons and time-varying trading costs, as well as clientele effects, etc. Amihud and Mendelson (1986) argue that the return of a stock increases with regard to its illiquidity, and that, in equilibrium, liquid assets with lower transaction costs are more likely to be held by high-frequency traders while the illiquid assets with higher transaction costs are to be held by long-term investors. Huang (2003) shows that in equilibrium, an illiquid stock with stochastic net return has a liquidity premium (over a liquid stock) that exceeds the expected transaction costs. Vayanos (2004) shows that liquidity premium is time-varying if an investor's liquidation risk (when to liquidate) is time-varying. In many cases, trading horizons are determined by investors' behavior, and they can tradeoff costs and benefits of delaying trades. Constantinides (1986) and Vayanos (1998) find that, in certain cases, liquidity premium is much smaller than trading costs. Vayanos (1998) also shows in certain cases, transaction costs actually can make stock price higher, as trading costs induce investors to hold shares longer, which in turn raises their demand.

**2.2 Liquidity Volatility Models**

Time-varying liquidity affects volatility as it affects price (Amihud, Mendelson and Pedersen, 2005). This suggests a positive relation between returns and volatility of liquidity. If a stock's liquidity has a high volatility, there is a high probability of low liquidity at liquidation. Therefore, risk-averse investors are likely to hold stocks with low volatility of liquidity. Petkova, Akbas and Armstrong (2011) find that stocks with higher variability in liquidity command higher expected returns, suggesting that investors dislike volatility of liquidity due to possible liquidity downturns. On the other hand, Chordia, Subrahmanyam and Anshuman (2001) show that there is a strong and negative relation between the volatility of liquidity and expected returns. Pereira and Zhang (2010) find that stocks with higher volatility of liquidity tend to earn lower returns, and that investors with investment horizons sell their stocks during periods of high liquidity, revealing their preference due to possible liquidity upturns. Amiram et al. (2019) consolidate the abovementioned dispute by decomposing the total volatility into a jump component and a diffusion component. They observe a positive relation between the jump volatility and illiquidity and a negative relation between the diffusive volatility and illiquidity, which translates to different effects on liquidity risk premium for the two volatility components.

Andersen, Bollerslev, and Meddahie (2011) extend existing diffusive volatility models to the market microstructure noise, in which liquidity is factored in implicitly. Their results suggest that the detrimental impact of the microstructure noise on forecast accuracy can be substantial. Bollerslev and Todorov (2023) argue that the risk associated with price jumps is priced differently from that with continuous price moves. They find that simultaneous jumps in the price and the stochastic volatility, and/or jump intensity of the market command a sizeable risk premium. Hu and Liu (2022) analyze the cross-section of index option returns by including volatility and jump risks, and argue that low option returns are primarily due to the pricing of volatility risk. Hussain (2011) uses an EGARCH model to estimate the conditional return volatility. The model includes contemporaneous and lagged trading volume and bid-ask spreads for information arrival, and tests asymmetric reactions of return volatility in response to changes in volume and spreads.

## 2.3 ARMA-GARCH Models and Volatility Dynamics

The ARMA-GARCH model capture conditional mean dynamics (via ARMA) and volatility clustering (via GARCH). However, their limitations in handling discontinuous jumps and liquidity-driven shocks have been well-documented. To address price jumps, researchers have extended ARMA-GARCH models. Chan and Maheu (2002) and Duan et al. (2006) introduce time-varying jump intensities and incorporate jump components to account for sudden price movements. Eraker, Johannes and Polson (2003) incorporate jumps in both returns and volatility. However, these models do not distinguish between jump sources (e.g., liquidity shocks vs. news events) but treat all jumps as homogeneous, ignoring their underlying causes. Bollerslev and Todorov (2011) emphasize the importance of extreme events but does not link jumps to liquidity. Similarly, Aït-Sahalia (2004) develops methods to disentangle jumps from continuous volatility but did not apply this to liquidity-driven discontinuities. These gaps are particularly problematic for cryptocurrencies, where jumps are often liquidity-induced (large trades on illiquid exchanges).

The exclusion of liquidity volatility is consequential, as liquidity fluctuations, particularly for assets traded in fragmented or illiquid markets, can amplify volatility in ways that traditional ARMA-GARCH models fail to capture (Brunnermeier and Pedersen, 2005). For example, unlike traditional assets such as stocks, cryptocurrencies trade 24/7 across many exchanges, mostly unregulated, with varying liquidity conditions (Dyhrberg, 2016). Baur and Dimpfl (2019) highlight that the volatility of Bitcoin (BTC) is more sensitive to liquidity shocks than equities, while Cong et al. (2023) show that wash trading distorts liquidity metrics in crypto markets. These findings underscore the inadequacy of traditional ARMA-GARCH models, which assume efficient and liquid markets. Furthermore, the interaction between liquidity and volatility is underexplored by the ARMA-GARCH framework. For cryptocurrencies, liquidity is inherently unstable due to factors like fragmented exchanges, thin order books, and the absence of market makers (Makarov and Schoar, 2020). These characteristics create feedback loops: declining liquidity exacerbates volatility, which further deters market participation (Bouchaud et al., 2018). However, traditional

ARMA-GARCH models do not explicitly model liquidity-volatility interactions, which may lead to biased parameter estimates for illiquid assets.

**2.4 Research Gaps**

At both theoretical and empirical levels, the literature research has yet developed a unified model that explicitly incorporates liquidity risk into return and volatility estimation. Although there exist studies on how liquidity interacts with return and volatility for both traditional assets (e.g., Amihud, Mendelson and Pedersen, 2005; Andersen, Bollerslev, and Meddahie, 2011; Amiram et al., 2019) and alternative assets (e.g., Manahov, 2021; Shen, Urquhart, and Wang, 2022), no attempts have been made to directly integrate liquidity measures to return and volatility so that liquidity risk can be modelled simultaneously and synchronously with return and volatility. We aim to bridge this gap by developing models that directly adjust return and volatility with liquidity measures, ensuring that liquidity risk is explicitly accounted for within the modeling process. The key benefit of our approach is that it automatically distinguishes jumps caused by liquidity shocks (already built-in to return and volatility) from other sources, enabling a more precise assessment of liquidity risk exposure.

At the methodological level, while the literature offers limited autoregressive models that address the liquidity risk (e.g., Hussain, 2011), these models are overly generic in scope (i.e., not asset specific) and rarely applied in asset pricing, as they involve cumbersome parameterization process. While there are autoregressive models for assets with extreme liquidity variability (e.g., Cortez, Rodríguez-García and Mongrutet, 2021), they tend to be "standard" models with no liquidity adjustment, and therefore not particularly effective for assets with high liquidity shocks. Some other studies model the liquidity effect implicitly through jumps with statistical models such as HAR (e.g., Corsi, 2009; Liu, Liu and Zhang, 2019), which essentially do not address the interaction between liquidity and volatility. In this paper, we present a new liquidity-adjusted ARMA-GARCH framework that delineate liquidity and volatility and address their interactions, which in turn reduces the biases in parameter estimation and therefore is particularly useful in modeling assets with low liquidity and high liquidity variability such as cryptocurrencies.

## 3. Liquidity Adjustment and ARMA-GARCH Models

### 3.1 Liquidity-adjusted Return and Liquidity-adjusted Volatility – minute-level

Imagine that a risk-averse investor had a choice between two assets, A and B, with identical expected return and volatility, yet A had a much higher trading volume/amount than B, the investor most likely would choose A, particularly if the investor's horizon is not indefinite. The intuitive reason for this choice is that the "perceived" volatility of A is lower than that of B. This choice is consistent with Amihud, Mendelson and Pedersen (2005). Using the minute-level trading data (crypto trading data from the largest cryptocurrency exchange Binance API, and stock trading data from the Polygon.io API), we model this perceived and unobservable volatility as a minute-level liquidity-adjusted variance $\sigma^2{}_T^\ell$ at equilibrium for time-period $T$ (a 24-hour/1440-minute trading day for cryptocurrencies and a 6.5-hour/390-minute trading day for the US stocks), with the variance of each minute being adjusted by an illiquidity factor similar to that of Amihud (2002). The expression of $\sigma^2{}_T^\ell$ is given as follows (using $\tau$ as minute-level time index for minute $\tau$):

$$\sigma^2{}_T^\ell = \frac{1}{T}\sum_{\tau=1}^T \eta_T \frac{|r_\tau|/\overline{|r_\tau|}}{A_\tau/\overline{A_\tau}} (r_\tau - \overline{r_\tau})^2 \Rightarrow \sigma^2{}_T^\ell = \frac{1}{T}\sum_{\tau=1}^T \left(\sqrt{\eta_T \frac{|r_\tau|/\overline{|r_\tau|}}{A_\tau/\overline{A_\tau}}} r_\tau - \sqrt{\eta_T \frac{|r_\tau|/\overline{|r_\tau|}}{A_\tau/\overline{A_\tau}}} \overline{r_\tau}\right)^2 \qquad (1)$$

$$\text{subject to: } \sum_{\tau=1}^T \eta_T \frac{|r_\tau|/\overline{|r_\tau|}}{A_\tau/\overline{A_\tau}} = T \Rightarrow \eta_T = \frac{T}{\sum_{\tau=1}^T \frac{|r_\tau|/\overline{|r_\tau|}}{A_\tau/\overline{A_\tau}}}$$

*where $r_\tau$ is the observed return at minute $\tau$, $|r_\tau|$ is its absolute value, $\overline{|r_\tau|}$ is its arithmetic average in that day,*
*$A_\tau$ is the dollar amount traded at minute $\tau$, $\overline{A_\tau}$ is its arithmetic average in day $T$,*
*$\eta_T$ is the daily normalization factor on day $T$ and is a constant for day $T$,*
*$T=1440$ or $390$ as there are 1440 minutes (24 hours) or 390 minutes (6.5 hours) in a crypto or US stock trading day*

Furthermore, in Equation 1, $\sigma^{2\ell}_T$ is the variance of an unobservable minute-level liquidity-adjusted return at equilibrium, $r^\ell_\tau$:

$$\sigma^{2\ell}_T = \frac{1}{T}\sum_{\tau=1}^{T}\left(r^\ell_\tau - \overline{r^\ell_\tau}\right)^2 \tag{2}$$

<small>where $r^\ell_\tau$ is the liquidity-adjusted return at minute $\tau$, $\overline{r^\ell_\tau}$ is its arithmetic average in day T.</small>

By equating the right-hand side of Equations 1 and 2, we model an asset's observed return $r_t$ and its unobservable liquidity-adjusted counterpart $r^\ell_\tau$ (1st order approximation) as follows[3]:

$$r^\ell_\tau = \sqrt{\eta_T \frac{|r_\tau|/|\overline{r_\tau}|}{A_\tau/\overline{A_\tau}}} r_\tau \tag{3}$$

<small>where $r_\tau$ is observed return at minute $\tau$, $r^\ell_\tau$ is the liquidity-adjusted return at minute $\tau$,<br>
$\beta^\ell_\tau$ is the liquidity factor at minute $\tau$.</small>

### 3.2 Liquidity-adjusted Return and Volatility, and Liquidity Beta Pair – daily-level

We then construct the unobservable daily-level liquidity-adjusted return by aggregating the intraday (minute-level) liquidity-adjusted returns. The realized and unobservable daily liquidity-adjusted returns on day $t$ are given as (using $t$ as time index for day $t$):

$$r_t = (1 + r_\tau)^T - 1 \tag{4a}$$
$$r^\ell_t = \left(1 + r^\ell_\tau\right)^T - 1 \tag{4b}$$

The realized and unobservable daily intraday (minute-level) variance for time-period $T$ is:

$$\sigma^{2\ell}_t = T\sigma^{2\ell}_T \tag{5}$$

In capture the liquidity dynamics on daily basis, we define a "liquidity Beta pair" that includes two liquidity measures, that one indicates the level of liquidity fluctuation (size), and one reflects the volatility of liquidity. We define a "liquidity jump," $\beta^\ell_{r_t}$, which is the measure of daily liquidity fluctuation for day $t$, as follow[4]:

---

[3] Note that although we define $r^\ell_\tau = \sqrt{\eta_T \frac{|r_\tau|/|\overline{r_\tau}|}{A_\tau/\overline{A_\tau}}} r_\tau$, in general $\overline{r^\ell_\tau} = \sqrt{\eta_T \frac{|r_\tau|/|\overline{r_\tau}|}{A_\tau/\overline{A_\tau}}} \overline{r_\tau}$ does not hold. However, its 1st order approximation is acceptable.

[4] It is possible that $r_t/r^\ell_t$ is negative, therefore $\beta^\ell_{r_t}$, a positive value, is defined as $|r_t/r^\ell_t|$.

$$\beta_{r_t}^{\ell} = |r_t/r_t^{\ell}| \subset \begin{cases} > 1; \text{high daily liquidity fluctuation} \\ = 1; \text{equilibrium daily liquidity fluctuation} \\ < 1; \text{low daily liquidity fluctuation} \end{cases} \qquad (6)$$

We also define a "liquidity diffusion," $\beta_{\sigma_t}^{\ell}$, which is the measure of daily (intraday) liquidity volatility for day *t*, as follow:

$$\beta_{\sigma_t}^{\ell} = \sigma_t/\sigma_t^{\ell} \subset \begin{cases} > 1; \text{high daily liquidity volatility} \\ = 1; \text{equilibrium daily liquidity volatility} \\ < 1; \text{low daily liquidity volatility} \end{cases} \qquad (7)$$

The liquidity jump $\beta_{r_t}^{\ell}$ captures the sudden and discontinuous change of daily liquidity level of an asset. A low-liquidity asset (e.g., a cryptocurrency) would have a higher $\beta_{r_t}^{\ell}$ value, indicating its high liquidity fluctuation, while a high-liquidity asset (e.g., US stocks) would have a lower $\beta_{r_t}^{\ell}$ value. The liquidity diffusion $\beta_{\sigma_t}^{\ell}$ reflects the intraday volatility of liquidity conditions, that an asset with stable liquidity (trading volume), such as a US stock, would have a lower $\beta_{\sigma_t}^{\ell}$ value, while an asset with high liquidity volatility induced by manipulative trades, such as a cryptocurrency, would have a higher $\beta_{\sigma_t}^{\ell}$ value. As such, cryptocurrencies would have higher liquidity jump ($\beta_{r_t}^{\ell}$) and higher liquidity diffusion ($\beta_{\sigma_t}^{\ell}$) than stocks, which indicates that cryptocurrencies, an asset class with low liquidity, have high liquidity variability in general.

## 4. Data and Descriptive Statistics

### 4.1 Dataset

We use the US stock trading data to establish and assess the benchmark of the liquidity-adjusted ARMA-GARCH framework. We collect minute-level trading data of all 1,503 constituent stocks of the SP500 (large cap), SP400 (mid cap) and SP600 (small cap) indices from the Polygon.io API. For each index, we pick five largest stocks in terms of market cap, thus we select 15 stocks from the three indices.[5] All 15 stocks have at least ten years of complete historical data (July 28, 2014 to February 10, 2025 with 2,652 trading days). The selected stocks, in alphabetical order, are AAPL, AMZN, ATI, CMA, CRS, EME, GOOG, IBKR, LII, MLI, MSFT, NVDA, TPL, VFC and WSO. We construct and calculate minute-level return and variance, both regular and liquidity-adjusted. From the minute-level data we calculate the daily return and (intraday) volatility, both regular ($r_t$ and $\sigma_t$) and liquidity-adjusted ($r_t^\ell$ and $\sigma_t^\ell$), as well as the daily liquidity jump $\beta_{r_t}^\ell$ and daily liquidity diffusion $\beta_{\sigma_t}^\ell$.[6] We report the descriptive statistics of $r_t$, $r_t^\ell$ and $\beta_{r_t}^\ell$ for all 15 US stocks in Table 1, and that of $\sigma_t$, $\sigma_t^\ell$ and $\beta_{\sigma_t}^\ell$ in Table 2. We also show the histograms of $\beta_{r_t}^\ell$ and $\beta_{\sigma_t}^\ell$ for all stocks in Figures 1 and 2, respectively.

We then collect tick-level trading data of the top 15 non-stable-coin cryptocurrencies (prices measured with their trading pairs with Tether or USDT[7]) in market cap from the Binance API[8] with at least four years of complete historical data (October 15, 2020 to February 7, 2025 with 1,577 trading days). We then discard cryptocurrencies that are either platform gas/incentive tokens (e.g., BNB) or task facility/incentive tokens (e.g., LINK), of which liquidity provision is affected by non-trading activities and therefore distorted, and end up with 10 of them. The selected cryptocurrencies, in alphabetical order, are ADA, AVAX, BCH, BTC, ETC, ETH, LTC, SOL, UNI and XRP. We calculate the same daily measures for the cryptocurrencies as that for the US stocks. We report the descriptive statistics of $r_t$, $r_t^\ell$ and $\beta_{r_t}^\ell$, as well as that of $\sigma_t$, $\sigma_t^\ell$ and $\beta_{\sigma_t}^\ell$ for all 10 cryptocurrencies in Tables 3 and 4, respectively. We also show the histograms of $\beta_{r_t}^{\ell}$ and $\beta_{\sigma_t}^{\ell}$ for all cryptocurrencies in Figures 3 and 4, respectively.

---

[5] We obtain the market caps of all 1,503 stocks from the Yahoo Finance API on October 12, 2024 at the market close.
[6] We cap the value of both $\beta_{r_t}^\ell$ and $\beta_{\sigma_t}^\ell$ to 10 in order to avoid any outliers to bias our analysis.
[7] USDT is a "stable coin" pegged to the US Dollar, and for our purpose of portfolio construction it is regarded as a "risk-free" asset with a 0% interest rate.
[8] https://coinmarketcap.com, accessed on October 12, 2024.

**4.2 Descriptive Statistics and Comparisons between Stocks and Cryptocurrencies**

In this subsection, we specifically discuss the descriptive statistics of liquidity jump ($\beta_{r_t}^{\ell}$) and liquidity diffusion ($\beta_{\sigma_t}^{\ell}$), and their differences between US stocks and cryptocurrencies, as they are directly related to the performance of the regular and liquidity-adjusted ARMA-GARCH models for both asset classes.

Panel C of Table 1 summarizes the descriptive statistics of liquidity jump $\beta_{r_t}^{\ell}$ for all stocks. For $\beta_{r_t}^{\ell}$, the mean is between 0.44 (TPL) and 1.39 (GOOG) and the median ranges from 0.20 (TPL) to 0.87 (AMZN). The number of days with $\beta_{r_t}^{\ell} \geq 1$ spans from 187 or 7.05% (TPL) to 1,015 or 38.27% (AMZN). Panel C of Table 3 summarizes the descriptive statistics of liquidity jump $\beta_{r_t}^{\ell}$ for all cryptocurrencies. The mean is between 1.54 (BTC) and 1.99 (BCH, UNI) and the median is between 0.85 (XRP) and 1.09 (SOL). The number of days with $\beta_{r_t}^{\ell} \geq 1$ ranges from 652 or 41.34% (XRP) to 868 or 55.04% (SOL). Comparing these descriptive statistics, we observe that for mean, median, and the number of days with high liquidity jump ($\beta_{r_t}^{\ell} \geq 1$), the values are numerically higher for cryptocurrencies than for stocks by a wide margin. This observation indicates that in general, cryptocurrencies have higher liquidity fluctuation than stocks, which is consistent with literature, which is validated by the visualizations of Figure 1 (histograms of stocks) and Figure 3 (histograms of cryptocurrencies) that show the liquidity jump $\beta_{r_t}^{\ell}$ of cryptocurrencies is more positively-skewed with longer right tail (higher bar at $\beta_{r_t}^{\ell} = 10$) than that of stocks.

Panel C of Table 2 provides descriptive statistics of liquidity diffusion $\beta^{\ell}_{\sigma_t}$ for all stocks. For $\beta^{\ell}_{\sigma_t}$, the mean is between 0.22 (TPL) and 0.79 (AMZN) and the median ranges from 0.21 (TPL) to 0.80 (AMZN). The number of days with $\beta^{\ell}_{\sigma_t} \geq 1$ is between 0 (CRS, EME, MLI, TPL and WSO) and 29 or 1.09% (NVDA), suggesting that liquidity volatility is very low for stocks. Panels C of Table 4 summarize the descriptive statistics of liquidity diffusion $\beta^{\ell}_{\sigma_t}$. The mean is between 0.82 (BTC) and 1.15 (BCH, UNI) and the median between 0.82 (BTC) and 1.06 (BCH). The number of days with high $\beta^{\ell}_{\sigma_t}$ ($\geq 1$) ranges from a low of 15 or 0.95% (BTC) to a high of 968 or 61.38% (BCH). By comparing the descriptive statistics of liquidity diffusion $\beta^{\ell}_{\sigma_t}$ for stocks and that for cryptocurrencies, we observe that for mean and median, the values are numerically higher for cryptocurrencies than for stocks, indicating that cryptocurrencies have higher liquidity volatility than stocks, which is consistent with literature. This observation is validated by the visualizations of Figure 2 (histograms of stocks) and Figure 4 (histograms of cryptocurrencies) that show the liquidity diffusion $\beta^{\ell}_{\sigma_t}$ of cryptocurrencies is more positively-skewed with longer right tail (higher bar at $\beta^{\ell}_{\sigma_t} = 10$) than that of stocks (with visible exceptions of BTC and ETH). For the number of days with high liquidity jump ($\beta^{\ell}_{\sigma_t} \geq 1$), the values are numerically higher for cryptocurrencies than for stocks by a large margin, with a very noticeable exception of BTC, of which the value is 0.95% and is actually lower than that of NVDA at 1.09%.

Combining the comparisons of the descriptive statistics of $\beta^{\ell}_{r_t}$ and $\beta^{\ell}_{\sigma_t}$, we establish that the US stocks display a pattern of very low level of liquidity diffusion (lower mean and media of $\beta^{\ell}_{\sigma_t}$, number of days with $\beta^{\ell}_{\sigma_t} \geq 1$) and reasonable low level of liquidity jump (lower mean and median, and number of days with $\beta^{\ell}_{r_t} \geq 1$) on daily level. On the other hand, cryptocurrencies display a pattern of high level of liquidity diffusion (higher mean and median of $\beta^{\ell}_{\sigma_t}$, and number of days with $\beta^\ell_{\sigma_t} \geq 1$) and high liquidity jump (higher mean and median of $\beta^\ell_{r_t}$, and higher number of days with $\beta^\ell_{r_t} \geq 1$). The differences of $\beta^\ell_{r_t}$ and $\beta^\ell_{\sigma_t}$ between stocks and cryptocurrencies are statistically significance based on a series of ANOVA tests.[9] That both liquidity jump and liquidity diffusion are higher for cryptocurrencies than stocks indicates that the stochastic jump innovation is more severe in the return for the former, and the liquidity adjustment may help remove the stochastic jump innovation from the return, restoring its responsiveness to the autoregressive model. This prompts us to propose that the liquidity-adjusted ARMA-GARCH model would have more observable performance improvement for cryptocurrencies than stocks. In the next section (Section 5) we provide detailed analyses and discussions on this proposition.

## 5. Liquidity-Adjusted ARMA-GARCH Model

In this section we discuss the liquidity-adjusted ARMA-GARCH framework for the modeling of univariate condition return and volatility for individual assets. A set of Adam-Fuller tests confirms that both $r_t$ and $r^\ell_t$ for all stocks and cryptocurrencies are stationary time series over the entire sample period for both stocks and cryptocurrencies.[10] We then apply the univariate ARMA($p,q$)-GARCH(1,1) construct to model both the regular return $r_t$ and liquidity-adjusted return $r^\ell_t$ for each asset in order to assess the improvement of model fit and the changes of key parameters, if any, from the regular model to the liquidity-adjusted model.

### 5.1 Liquidity-adjusted ARMA-GARCH Model

---

[9] We do not report the results of these ANOVA tests, as comparing $\beta^\ell_{r_t}$ and $\beta^\ell_{\sigma_t}$ is not the focus for this paper. These results are available upon request.
[10] We do not report the results of these Adam-Fuller tests in the paper for the purpose of being concise. These results are available upon request.

The standard univariate ARMA($p,q$)-GARCH(1,1) specification that models the regular daily return $r_t$ and conditional volatility $\sigma_t^\omega$ (not to be confused with intraday volatility $\sigma_t$) is given as:

$$ARMA(p,q): r_t = \sum_{i=1}^{p} \phi_i r_{t-i} + \sum_{j=1}^{q} \theta_j \epsilon_{t-j} + \epsilon_t = \mu_t + \epsilon_t \quad (8)$$

$$GARCH(1,1): \sigma_t^{\omega 2} = \omega + a\epsilon_{t-1}^2 + b\sigma_{t-1}^{\omega 2}$$

where:

$$\mu_t = \sum_{i=1}^{p} \phi_i r_{t-i} + \sum_{j=1}^{q} \theta_j \epsilon_{t-j}$$

$$\epsilon_t | \Psi_{t-1} = \sigma_t^\omega z_t; z_t \sim \mathcal{N}(0,1)$$

In Equation 8, $\mu_t$ is the expected return of $r_t$. Furthermore, without liquidity adjustment to, the unincorporated information on liquidity risk can be modeled as a stochastic jump innovation ($\epsilon_t^J$) in the ARMA($p,q$)-GARCH(1,1) specification as:

$$ARMA(p,q): r_t = \sum_{i=1}^{p} \phi_i r_{t-i} + \sum_{j=1}^{q} \theta_j \epsilon_{t-j} + \epsilon_t + \epsilon_t^J = \mu_t + \epsilon_t + \epsilon_t^J \quad (9)$$

$$GARCH(1,1): \sigma_t^{\omega 2} = \omega + a\epsilon_{t-1}^2 + b\sigma_{t-1}^{\omega 2}$$

where:

$$\mu_t = \sum_{i=1}^{p} \phi_i r_{t-i} + \sum_{j=1}^{q} \theta_j \epsilon_{t-j}$$

$$\epsilon_t | \Psi_{t-1} = \sigma_t^\omega z_t; z_t \sim \mathcal{N}(0,1)$$

$\epsilon_t^J$ is the jump innovation

In literature, the stochastic jump innovation $\epsilon_t^J$ in ARMA-GARCH is typically modeled (e.g. Zhu, Song and Zheng, 2024) as a collection of $n_t$ number of jumps arriving in the time interval $[t-1, t]$ (in the context of this paper, $n_t$ number of minute-level jumps in day $t$). The stochastic variable $n_t$ follows a Poisson distribution with intensity $P(n_t = j)$ in day $t$ on the minute-level, and the size of each minute-level jump $Y_t^\tau$ follows a normal distribution. Therefore, $\epsilon_t^J$ is given as:

$$\epsilon_t^J = \sum_{\tau=1}^{n_t} Y_t^\tau \quad (10)$$

where:

$$P(n_t = j) = \frac{\lambda_t^j e^{-\lambda_t}}{j!}$$

$$Y_t^\tau \sim \mathcal{N}\left(\overline{Y_t^\tau}, \sigma_t(Y_t^\tau)\right)$$

In Equation 10, $\epsilon_t^J$ is the daily-level aggregation of $n_t$ minute-level innovations induced by extreme liquidity, thus $n_t$ is less than 390 for stocks or 1,440 for cryptocurrencies ($t$ is the daily-level time index, $\tau$ is the minute-level time index with a value up to $n_t$). As $\beta_{r_t}^\ell$ is the aggregated daily-level liquidity jump, it can be used to measure the size of $\epsilon_t^J$ ($\sum_{\tau=1}^{n_t} Y_t^\tau$); also as $\beta_{\sigma_t}^\ell$ is the daily intraday liquidity volatility (intensity), it can be used to measure the jump arrival intensity of $\epsilon_t^J$ ($\frac{\lambda_t^j e^{-\lambda_t}}{j!}$). Therefore, $\epsilon_t^J$ is the daily-level aggregation of minute-level stochastic jump innovations. It is a function of both $\beta_{r_t}^\ell$ and $\beta_{\sigma_t}^\ell$, and linear to the expected return $\mu_t$, which is expressed as:

$$\epsilon_t^J = \gamma_t(\beta_{r_t}^\ell) \lambda_t(\beta_{\sigma_t}^\ell) \mu_t \tag{11}$$

$$\Rightarrow \epsilon_t^J = (\beta_{r_t}^\ell - 1)\beta_{\sigma_t}^\ell \mu_t \tag{12}$$

In Equation 11, $\gamma_t(\beta_{r_t}^\ell)$ and $\lambda_t(\beta_{\sigma_t}^\ell)$ are the size and intensity of liquidity-induced jump on day $t$, respectively, and Equation 12 is one simple representation of Equation 11. We recognize that in Equation 12, $\epsilon_t^J$ is no longer stochastic, which is due to that $\epsilon_t^J$ is the deterministic daily-level manifestation of aggregated minute-level stochastic jump innovations. We then substitute the $\epsilon_t^J$ term in Equation 9 with Equation 12 and get:

$$r_t = \sum_{i=1}^p \phi_i r_{t-i} + \sum_{j=1}^q \theta_j \epsilon_{t-j} + \epsilon_t + \epsilon_t^J = \mu_t + \epsilon_t + \epsilon_t^J$$
$$\Rightarrow r_t = \mu_t + \epsilon_t + (\beta_{r_t}^\ell - 1)\beta_{\sigma_t}^\ell \mu_t$$
$$\Rightarrow r_t = [1 + (\beta_{r_t}^\ell - 1)\beta_{\sigma_t}^\ell]\mu_t + \epsilon_t \tag{13}$$

In Equation 13, as $\epsilon_t^J$ is no longer stochastic, we incorporate it to the new expected return, which now has a scaling coefficient.

$$\Rightarrow \frac{r_t}{\beta_{r_t}^\ell} = \frac{1 + (\beta_{r_t}^\ell - 1)\beta_{\sigma_t}^\ell}{\beta_{r_t}^\ell} \mu_t + \frac{\epsilon_t}{\beta_{r_t}^\ell}$$
$$\Rightarrow r_t^\ell = \mu_t^\ell + \epsilon_t^\ell \tag{14}$$

where:

$$r_t^\ell = \frac{r_t}{\beta_{r_t}^\ell}; \quad \epsilon_t^\ell = \frac{\epsilon_t}{\beta_{r_t}^\ell}$$

$$\mu_t^\ell = \frac{1+(\beta_{r_t}^\ell-1)\beta_{\sigma_t}^\ell}{\beta_{r_t}^\ell}\mu_t$$

Equation 14 is essentially an ARMA model of liquidity-adjusted return $r_t^\ell$, with $\mu_t^\ell$ being the liquidity-adjusted expected return. We derive the distribution of $\epsilon_t^\ell$ from Equation 9 as:

$$\epsilon_t|\Psi_{t-1} = \sigma_t^\omega z_t; z_t \sim \mathcal{N}(0,1)$$
$$\Rightarrow \frac{\epsilon_t}{\beta_{r_t}^\ell}|\Psi_{t-1} = \frac{\sigma_t^\omega}{\beta_{r_t}^\ell}z_t; z_t \sim \mathcal{N}(0,1)$$
$$\Rightarrow \epsilon_t^\ell|\Psi_{t-1} = \sigma_t^{\omega\ell} z_t; z_t \sim \mathcal{N}(0,1) \qquad (15)$$
where:
$$\epsilon_t^\ell = \frac{\epsilon_t}{\beta_{r_t}^\ell}; \sigma_t^{\omega\ell} = \frac{\sigma_t^\omega}{\beta_{r_t}^\ell}$$

Equation 15 shows that the liquidity-adjusted error $\epsilon_t^\ell$ is equal to $\epsilon_t$ being scaled by $\beta_{r_t}^\ell$, and that the conditional volatility $\sigma_t^{\omega\ell}$ is equal to $\sigma_t^\omega$ being scaled by $\beta_{r_t}^\ell$. As such, when $\beta_{r_t}^\ell > 1$, $\epsilon_t^\ell < \epsilon_t$ and $\sigma_t^{\omega\ell} < \sigma_t^\omega$; when $\beta_{r_t}^\ell < 1$, $\epsilon_t^\ell > \epsilon_t$ and $\sigma_t^{\omega\ell} > \sigma_t^\omega$. Numerically, this means that when liquidity jump is high ($\beta_{r_t}^\ell > 1$), the ARMA-GARCH model produces lower error term and lower conditional volatility for the liquidity-adjusted return $r_t^\ell$ than for the regular return $r_t$, and when liquidity jump is low ($\beta_{r_t}^\ell < 1$) the opposite is true. Conceptually, this indicates that liquidity adjustment compensates volatility swings.

As presented in Subsection 4.2, for all stocks, the mean of $\beta_{r_t}^\ell$, is between 0.44 (TPL) and 1.39 (GOOG) and the number of days with $\beta_{r_t}^\ell \geq 1$ spans from 187 or 7.05% (TPL) to 1,015 or 38.27% (AMZN) (Panel C of Table 1). This means that the number of days with $\beta_{r_t}^\ell < 1$ is between 61.73% and 92.95%, which essentially means that the error and conditional volatility of the liquidity-adjusted model are higher than that of the traditional model (i.e., $\epsilon_t^\ell > \epsilon_t$ and $\sigma_t^{\omega\ell} > \sigma_t^\omega$) for a majority of trading days. Therefore, we expect that the liquidity-adjusted ARMA-GARCH model does not provide a model fit improvement over the traditional model for stocks.

On the other hand, the number of days with $\beta_{r_t}^\ell \geq 1$ ranges from 652 or 41.34% (XRP) to 868 or 55.04% (SOL), thus the number of days with $\beta_{r_t}^\ell < 1$ is between 44.96.73% and 58.66%, and therefore there are about an even split of days with $\epsilon_t^\ell > \epsilon_t$ and $\sigma_t^{\omega^\ell} > \sigma_t^\omega$ and $\epsilon_t^\ell < \epsilon_t$ and $\sigma_t^{\omega^\ell} < \sigma_t^\omega$. Combining this with the fact that the mean of $\beta_{r_t}^\ell$ for all cryptocurrencies is between 1.54 (BTC) and 1.99 (BCH, UNI), all greater than 1 by a sizable margin, we anticipate that the liquidity-adjusted ARMA-GARCH model should improve the model fit over the traditional model for a majority of cryptocurrencies. This is the key methodological reason why we predict that the liquidity-adjusted ARMA-GARCH model would improve upon the traditional ARMA-GARCH model more for cryptocurrencies than stocks.

Although Equations 14 and 15 can be indeed used as a liquidity-adjusted alternative to the standard ARMA-GARCH model of Equation 9 (with jump), their calculation is non-trivial and involves multiple recursive steps. If, however, we use ARMA-GARCH to model the liquidity-adjusted return directly, we can take advantage of the simple ARMA-GARCH construct, and rewrite Equations 14 and 15 with a liquidity-adjusted ARMA($p,q$)-GARCH(1,1) specification as:

$$r_t^\ell = \sum_{i=1}^p \phi_i^\ell r_{t-i}^\ell + \sum_{j=1}^q \theta_j^\ell \epsilon_{t-j}^\ell + \epsilon_t^\ell = \mu_t^\ell + \epsilon_t^\ell \tag{16}$$

$$\sigma_t^{\omega^\ell 2} = \omega^\ell + a^\ell {\epsilon_{t-1}^\ell}^2 + b^\ell \sigma_{t-1}^{\omega^\ell 2}$$

$$\text{where}:$$

$$\mu_t^\ell = \sum_{i=1}^p \phi_i^\ell r_{t-i}^\ell + \sum_{j=1}^q \theta_j^\ell \epsilon_{t-j}^\ell$$

$$\epsilon_t^\ell | \Psi_{t-1} = \sigma_t^{\omega^\ell} z_t; z_t \sim \mathcal{N}(0,1)$$

In Equation 16, in general, the ARMA parameters ($p, q$) are not equal to the ARMA parameters ($p, q$) in Equation 9, and therefore $\mu_t^\ell$, $\epsilon_t^\ell$ and conditional volatility $\sigma_t^{\omega^\ell}$ are not equal to their corresponding values in Equations 14 and 15. However, the expectation that the liquidity-adjusted ARMA-GARCH model would improve upon the traditional ARMA-GARCH model more for cryptocurrencies than stocks still holds. As such, we transfer the regular ARMA-GARCH with stochastic jump to a liquidity-adjusted ARMA-GARCH without.

**5.2 Modeling for Stocks**

We first compare the model fit and key parameter estimates between the regular and liquidity-adjusted ARMA-GARCH models for US stocks for in-sample observations with a 242-day rolling window, covering the period from July 14, 2015 to February 10, 2025, with 2,410 trading days.

In the ARMA stage, we use the Akaike Information Criterion (AIC) to determine the optimal values of $p$ and $q$ ($p,q \leq 4$) for each rolling window. In the GARCH stage, we adopt the GARCH(1,1) specification to reduce computation penalty for each in-sample day $t$. To assess model performance, we compare the log-likelihood values of the standard ARMA-GARCH model against those of the liquidity-adjusted ARMA-GARCH model using a pairwise t-test with three alternative hypotheses: two-sided, less, and greater. The results for each stock are presented in Panel A of Table 5.[11]

The findings indicate that for all US stocks analyzed, the standard ARMA-GARCH model exhibits a statistically superior model fit compared to the liquidity-adjusted version at the 1% significance level. This result aligns with our expectations, as liquidity risk is less critical for highly liquid assets like US stocks. The implication is that the regular return of a US stock already has an adequate amount of liquidity information implicitly priced in by the market, and therefore the liquidity adjustment does not add a statistically significant amount of extra liquidity information to the return. Consequently, for US stocks, the liquidity-adjusted ARMA-GARCH model does not provide a better perspective on volatility than the regular model.

---

[11] We do not report the test results with the AIC and BIC criteria as they yield the same model fit for each of the stocks. The test results with the AIC and BIC criteria are available upon request.

The decline in model fit from the regular ARMA-GARCH model to the liquidity-adjusted model can be attributed to the changes of key parameters in the ARMA-GARCH model, particularly the two coefficients in the GARCH(1,1) specification: $a$ and $b$ in Equation 8 for the regular return, and $a^\ell$ and $b^\ell$ in Equation 16 for the liquidity-adjusted return. The "shock coefficients" of $a$ and $a^\ell$ quantifies the sensitivity of current volatility to past shocks, with a higher value indicating stronger sensitivity to recent shocks. The "volatility coefficients" of $b$ and $b^\ell$ measure how long past volatility effects persist over time, where a higher value signifies stronger autocorrelation in volatility.

Panel B of Table 5 compares the shock coefficients of $a$ for the regular model and $a^\ell$ for the liquidity-adjusted model. Out of the 15 stocks, 11 stocks have their $a$ value reduced to a lower $a^\ell$ value, two (CRS, MSFT) have their $a$ value increased to a higher $a^\ell$ value, and two (AMZN, NVDA) do not have statistically different $a$ and $a^\ell$ values. The results indicate that the liquidity adjustment reduces the sensitivity of volatility to past shocks for a large majority of selected stocks. As we establish that the regular return of a stock, a highly liquid asset with sufficient liquidity information already priced in, these results suggest that the impact of liquidity risk embedded in the past shocks may have been appropriately estimated by the ARMA-GARCH model, and the liquidity adjustment actually removes some of the liquidity information on past shocks, which in turn degrades the model fit. Still, there is a small number of stocks with increased sensitivity of volatility to past shocks, therefore the above assessment is not conclusive for all stocks.

Panel B of Table 5 presents a comparison of shock coefficients between the standard and liquidity-adjusted models. Among the 15 stocks analyzed, 11 stocks exhibit a decrease in their shock coefficient after liquidity adjustment ($a > a^{\ell}$), while two stocks (CRS, MSFT) experience an increase ($a < a^{\ell}$). The remaining two stocks (AMZN, NVDA) do not show statistically significant difference between $a$ and $a^{\ell}$ values. These results suggest that in most cases, liquidity adjustment reduces the sensitivity of volatility to past shocks. Given that the regular return of a stock, a highly liquid asset, already have sufficient liquidity information priced in, these results imply that the impact of liquidity risk embedded in the past shocks may have been appropriately estimated by the ARMA-GARCH model, and the liquidity adjustment may actually remove some of the liquidity information, which degrades model fit. However, since two stocks show an increase in shock sensitivity after liquidity adjustment, the above argument does not hold universally across all stocks.

Panel C of Table 5 compares the volatility coefficients $b$ for the regular ARMA-GARCH return and $b^{\ell}$ for the liquidity-adjusted model. Out of the 15 stocks, eight have their volatility coefficients increased ($b < b^{\ell}$), while five have them reduced ($b > b^{\ell}$), and two (EME, MLI) do not experience statistically different $b$ and $b^{\ell}$ values. These findings suggest that for a slight majority of stocks, liquidity adjustment increases the persistence of volatility, while for a smaller subset, it decreases persistence. Given this mixed evidence, no clear directional effect of liquidity adjustment on volatility persistence for US stocks as an asset class can be determined.

Overall, the analysis reveals that the liquidity-adjusted ARMA-GARCH model does not offer superior model fit over the standard model for US stocks, largely because these stocks already incorporate sufficient liquidity information in their return dynamics. The liquidity adjustment tends to reduce the sensitivity of volatility to past shocks for most stocks while yielding inconclusive effects on volatility persistence. This outcome aligns with the expectation that exposure to liquidity risk is relatively how for highly liquid assets, and traditional ARMA-GARCH modeling is sufficient for modeling US stocks without liquidity adjustment.

**5.3 Modeling for Cryptocurrencies**

We then compare the model fit and key parameter estimates between the regular and liquidity-adjusted ARMA-GARCH models for cryptocurrencies, using in-sample observations with a 365-day rolling window (from October 15, 2021 to February 7, 2025 with 1,212 trading days).

In Panel A of Table 6, the results show that, out of the 10 selected cryptocurrencies, seven have their model fit improved by the liquidity-adjusted model at 1% significant level, and two experience no statistically significant improvement (although they both have a negative $t$-value, indicating a non-significant improvement). The only cryptocurrency with a declined model fit is BTC. The improvement of model fit for cryptocurrencies is quite the opposite to that for stocks, with a very noticeable exception of BTC. This pattern contrasts sharply with the results for stocks. The findings suggest that, in general, the regular return of a cryptocurrency, an asset with low liquidity and high liquidity variability, may not have sufficient amount of liquidity information priced in by the market, and that the liquidity adjustment indeed supplements a statistically significant amount of additional liquidity information to the return. As such, the liquidity-adjusted ARMA-GARCH model provides better perspective on volatility than the regular model. However, BTC, as the largest cryptocurrency in both market cap and liquidity, has a liquidity profile that is similar to that of stocks, and therefore just like stocks, the liquidity adjustment does not benefit from liquidity adjustment, as its return dynamics already reflect sufficient liquidity information.

Panel B of Table 6 compares the shock coefficients of $a$ for the regular model and $a^\ell$ for the liquidity-adjusted model for the selected cryptocurrencies. Among the 10 cryptocurrencies, nine exhibit an increase in shock coefficient ($a < a^{\ell}$), and one (ETC) does not have statistically different $a$ and $a^{\ell}$ values (although *t*-value is negative, indicating a non-significant increase). The is again the opposite to the case for stocks. The results show that the liquidity adjustment enhances the sensitivity of volatility to past shocks for essentially all selected cryptocurrencies. A likely explanation is that the frequent sudden liquidity shocks experienced by cryptocurrencies may have been underestimated by the traditional ARMA-GARCH model, and the liquidity adjustment improves the sensitivity of volatility to these sudden liquidity shocks. The improved sensitivity to volatility is one of the primary reasons why the liquidity adjustment improves the ARMA-GARCH model fit for cryptocurrencies.

Panel C of Table 6 outlines the comparisons between volatility coefficients $b$ for the regular return and $b^{\ell}$ for the liquidity-adjusted model. Out of the 10 cryptocurrencies, seven have their volatility coefficient reduced ($b > b^{\ell}$), while three (ADA, BCH, ETC) show no statistically significant difference between $b$ and $b^{\ell}$. Unlike in the case of stocks, where the impact of liquidity adjustment on volatility persistence is inconclusive, the findings for cryptocurrencies present a clearer pattern. The results suggest that liquidity adjustment reduces the persistence of volatility, making current volatility less dependent on erratic past volatility typical for cryptocurrencies. The reduced volatility persistence allows for "smoother" responses to the lingering effects of previous extreme fluctuations. This reduction in volatility persistence further explains why liquidity adjustment improves the ARMA-GARCH model fit for cryptocurrencies.

Overall, these results demonstrate that the liquidity-adjusted ARMA-GARCH model provides a significantly better fit for most cryptocurrencies, as it effectively accounts for the frequent liquidity shocks and high liquidity variability in cryptocurrency trading. The liquidity-adjusted model enhances the sensitivity of volatility to past shocks and reduces the persistence of volatility, leading to a more responsive and adaptive volatility estimation. The exception is BTC, which, due to its higher liquidity and market efficiency, does not benefit from liquidity adjustments, much like traditional stocks. These findings highlight the importance of liquidity-adjusted modeling for cryptocurrencies and suggest that the liquidity-adjusted ARMA-GARCH framework is more effective in capturing the volatility dynamics of low-liquidity and high-volatility assets.

## 6. Mean Variance (MV) Portfolio Optimization: Empirical Tests

In order to provide empirical evidence to support that the liquidity-adjusted ARMA-GARCH model indeed improves upon the traditional ARMA-GARCH with better model fit for cryptocurrencies, we apply the ARMA($p,q$)-GARCH(1,1) specification to forecast the one-period ($t+1$) forward conditional mean return ($\hat{\mu}_{r_{t+1}}^{arga}, \hat{\mu}_{r_{t+1}^{\ell}}^{arga}$) for each rolling window for both $r_t$ and $r_t^{\ell}$,[12] for each of the selected US stocks and cryptocurrencies. For the regular return $r_t$, we use $\hat{\mu}_{r_{t+1}}^{arga}$ as input to a traditional MV (TMV) portfolio. The portfolio consists of the returns of two assets, the regular return $r_t$ of either a stock or a cryptocurrency, and a risk-free asset (which we assume a 0% return to simplify analysis). As such, there are 15 TMV portfolios for stocks and 10 TMV portfolios for cryptocurrencies. Similarly, for the liquidity-adjusted return $r_t^{\ell}$, we apply $\hat{\mu}_{r_{t+1}^{\ell}}^{arga}$ as input to a liquidity-adjusted MV portfolio (LAMV) for each asset, and there are 15 and 10 LAMV portfolios for stocks and cryptocurrencies, respectively.

### 6.1 ARMA-GARCH-enhanced TMV and LAMV Portfolios

---

[12] For the rest of the paper, the "current timestamp" is end of day $t$, thus a variable with a $t$ subscript is "realized" (either a direct observation or a calculated value from direct observations), while a variable with a $t+1$ subscript is a one-period forecasted value. We also use accent mark "$\bar{v}$" to represent a mean variable, accent mark "$\hat{v}$" to represent a forecasted variable, and no accent mark "$v$" to represent a realized variable.

For each asset (either stock or cryptocurrency), we construct two MV portfolios: TMV and LAMV. The standard daily-optimized MV in a time-series construct can be analytically expressed as the following quadratic programming problem with constraints:

$$\max_{W_t} \left( \bar{\mu}_t W_t^{Tr} - \frac{\lambda_t}{2} W_t^{Tr} \bar{\Sigma}_t W_t \right); Tr \text{ stands for Transposed} \qquad (17)$$

*subject to:*

$w_t^i + w_t^{r_f} = 1$; $i$ is the asset (stock or crypto), $r_f$ is risk-free asset

$w_t^i \geq 0, w_t^i \leq 0$; *long-only*

$w_t^{r_f} \leq 1$

*where:*

$$\lambda_t = \frac{r_{t_{mkt}}^P - r_{t_{rf}}}{\sigma_{t\,mkt}^2} = \frac{r_{t_{mkt}}^P}{\sigma_{t\,mkt}^2}$$

$r_{t_{mkt}}^P$ *is the return of the market portfolio on day t,* $\sigma_{t\,mkt}^2$ *is the variance;*
*the market portfolio: either 15 stocks for the stock market portfolio, or 10 cryptos for the cryto market portfolio*
$r_{t_{rf}}$ *is the return of a risk-free asset, and with 0% return for simplicity*

In the standard MV construct of Equation 17, $\bar{\mu}_t$ is the mean return vector of the two assets (1 stock or crypto + 1 risk-free asset such as cash) portfolio over a rolling window (242 days for stocks, 365 days for cryptos) ending on day $t$, and $\bar{\Sigma}_t$ is the covariance matrix of the constituent assets in that rolling window. Both $\bar{\mu}_t$ and $\bar{\Sigma}_t$ are derived from available information up to day $t$. In addition, $W_t$ is the portfolio weight vector to be optimized for day $t$. We rewrite Equation 17 by retaining $\bar{\Sigma}_t$ and replacing $\bar{\mu}_t$ with the ARMA-GARCH forecasted return vector ($\hat{\mu}_{t+1}^{arga}$) as:

$$\max_{W_t} \left( \hat{\mu}_{t+1}^{arga} W_t^{Tr} - \frac{\lambda_t}{2} W_t^{Tr} \bar{\Sigma}_t W_t \right) \qquad (18)$$

All constraints for Equation 18 are the same as that in Equation 17. The ARMA-GARCH-enhanced MV portfolios are:

1. Portfolio 1: The ARMA-GARCH-enhanced TMV portfolio; $\hat{\mu}_{t+1}^{arga}$ is the return vector of ARMA-GARCH forecasted $r_t$ values for day $t+1$, $\hat{\mu}_{r_{t+1}}^{arga}$; $\bar{\Sigma}_t$ is the covariance matrix of $r_t$'s for the rolling window, or $\bar{\Sigma}_{r_t}$.

2. Portfolio 2: The ARMA-GARCH-enhanced LAMV portfolio; $\hat{\mu}_{t+1}^{arga}$ is the return vector of ARMA-GARCH forecasted $r_t^\ell$ values for day $t+1$, $\hat{\mu}_{r_{t+1}^\ell}^{arga}$; $\bar{\Sigma}_t$ is the covariance matrix of $r_t^\ell$'s for the rolling window, or $\bar{\Sigma}_{r_t^\ell}$.

## 6.2 Empirical Tests: Performance Comparisons of TMV and LAMV Portfolios

We use the annualized Sharpe Ratio ($SR_a$) to compare the performance between portfolios:

$$SR_a = \frac{r_a^P - r_a^{rf}}{\sigma_a^P} = \frac{r_a^P}{\sigma_a^P} \tag{19}$$

Where:
1. $r_a^P, \sigma_a^P$ are the annualized realized regular daily portfolio return and standard deviation.
2. $r_a^{rf}$ is the annualized realized daily returns for the risk-free asset, and assumed to be 0 for simplicity.

Panel A of Table 7 presents the portfolio performance (Sharpe Ratio) for each of the 15 TMV and 15 LAMV portfolios for stocks. Among these, nine exhibit an improvement in SR by liquidity adjustment (Portfolio 2 has higher SR than Portfolio 1), while five have a decline in SR (Portfolio 2 has lower SR than Portfolio 1), and one has no statistically significant change of SR. These mixed results suggest that the effectiveness of liquidity adjustment on stock return depends largely on the extent of liquidity information already embedded in that particular return. However, the findings remain inconclusive on whether the liquidity adjustment improves the model fit of ARMA-GARCH for stocks in general.

Panel B of Table 7 provides the portfolio performance (Sharpe Ratio) for each of the 10 TMV and 10 LAMV portfolios for cryptocurrencies. With the exception of BTC, all cryptocurrencies show an improvement in SR with the liquidity adjustment (Portfolio 2 has higher SR than Portfolio 1). This strong and consistent empirical evidence confirms that the liquidity adjustment incorporates additional liquidity information into the return, which in turn improves the model fit of ARMA-GARCH for cryptocurrencies.

In summary, we find that the LAMV portfolio does not offer improved performance over the TMV portfolio for stocks, which validates our assessment that the liquidity-adjusted ARMA-GARCH does not offer better model fit for assets with high liquidity and low liquidity variability. On the other hand, we demonstrate that the LAMV portfolio has a clear performance advantage over the TMV portfolio for all cryptocurrencies except for BTC, which provides empirical evidence to our proposition that the liquidity-adjusted ARMA-GARCH offers better model fit for assets with low liquidity and high liquidity variability. In the special case of BTC, as the largest cryptocurrency in terms of both market cap and liquidity, it exhibits liquidity variability similar to stocks, therefore the liquidity-adjusted ARMA-GARCH model does not offer better model fit.

## 7. Conclusions

We show that the traditional ARMA-GARCH framework does not adequately capture the effect of liquidity on the returns of cryptocurrencies. The reason is that the returns of cryptocurrencies – as an asset class with low liquidity and high liquidity variability – may not have sufficient amount of liquidity information priced in by the market, leading to systematically biased estimates on model parameters. To address this limitation, we develop a pair of liquidity measures (liquidity jump and liquidity diffusion) and adjust the return and volatility dynamics with directly integrated liquidity information. We demonstrate that the liquidity adjustment indeed supplements a statistically significant amount of additional liquidity information to the returns, which renders the ARMA-GARCH effective again. As such, the liquidity-adjusted ARMA-GARCH framework provides better perspective on volatility than the regular model. In essence, the proposed model extends the traditional ARMA-GARCH to account for liquidity risk, resulting in more robust and responsive volatility forecasts for assets prone to liquidity shocks.

We provide analyses on U.S. stocks and cryptocurrencies, which validate the effectiveness of the liquidity-adjusted ARMA-GARCH framework. For assets with ample liquidity and stable liquidity variability (e.g., U.S. stocks), the liquidity-adjusted framework actually has degraded model fit than its traditional counterpart, suggesting that the returns of these assets already have adequate liquidity information priced in by the market, and the liquidity adjustment does not add a statistically significant amount of extra liquidity information to the returns. However, for illiquid assets with high liquidity variability (e.g., cryptocurrencies), the liquidity-adjusted ARMA-GARCH framework has a significantly improved model fit over the traditional model. This suggests that the returns of cryptocurrencies may not have sufficient amount of liquidity information priced in by the market, and that the liquidity adjustment indeed supplements a statistically significant amount of additional liquidity information to the returns. The largest cryptocurrency in terms of both market cap and liquidity, BTC, is a notable exception, as it has a liquidity profile that is actually similar to that of stocks, and therefore the liquidity adjustment does not improve its model fit.

In these cases, the enhanced model delivered more accurate volatility forecasts and improved predictability of returns. For instance, a mean-variance portfolio guided by the liquidity-adjusted model's forecasts (LAMV) consistently outperformed a similar portfolio based on the traditional model's forecasts (TMV) in the cryptocurrency market. This empirical evidence confirms that accounting for liquidity risk markedly improves forecasting and risk-adjusted performance for assets characterized by frequent liquidity shocks.

We provide empirical evidence that the liquidity-adjusted ARMA-GARCH framework indeed offers better model fit by comparing the performances of a series of two-asset LAMV and TMV portfolios. We demonstrate that the LAMV portfolio does not offer improved performance over the TMV portfolio for stocks, but has a clear performance advantage for all cryptocurrencies except for BTC. The empirical evidence supports our assessment that, while the liquidity-adjusted ARMA-GARCH framework does not offer better model fit for assets with high liquidity and low liquidity variability over the traditional model (i.e., stocks), it indeed offers better model fit for assets with low liquidity and high liquidity variability (i.e., cryptocurrencies).

In summary, our findings carry important implications for liquidity risk management. By incorporating an adequate amount of liquidity information dynamics explicitly in the turn and volatility dynamics of illiquid assets with high liquidity variability, the liquidity-adjusted ARMA-GARCH framework presented in this paper is a viable and robust alternative for jump and statistical models (e.g., HAR) and represents a significant step toward more accurate and liquidity-aware financial modeling, ultimately contributing to better performance and more robust portfolio management for assets subject to excessive liquidity risk.

# References

Acharya, V.V. and Pedersen, L.H., 2005. Asset pricing with liquidity risk. *Journal of Financial Economics*, 77(2), pp.375-410.

Aït-Sahalia, Y., 2004. Disentangling diffusion from jumps. *Journal of Financial Economics*, 74(3), pp.487-528.

Amihud, Y., 2002. Illiquidity and stock returns: cross-section and time-series effects. *Journal of Financial Markets*, 5(1), pp.31-56.

Amihud, Y. and Mendelson, H., 1986. Liquidity and stock returns. *Financial Analysts Journal*, 42(3), pp.43-48.

Amihud, Y., Mendelson, H. and Pedersen, L.H., 2005. Liquidity and asset prices. *Foundations and Trends® in Finance*, 1(4), pp.269-364.

Amiram, D., Cserna, B., Kalay, A. and Levy, A., 2019. The Information Environment, Volatility Structure, and Liquidity. *Columbia Business School Research Paper*, 15-62.

Anderson, T.G., Bollerslev, T. and Meddahi, N., 2011. Realized volatility forecasting and market microstructure noise. *Journal of Econometrics*, 160, pp.220-234.

Baur, D.G. and Dimpfl, T., 2019. Price discovery in bitcoin spot or futures? *Journal of Futures Markets*, 39(7), pp.803-817.

Bollerslev, T., 1986. Generalized autoregressive conditional heteroskedasticity. *Journal of Econometrics*, 31(3), pp.307-327.

Bollerslev, T. and Todorov, V., 2011. Estimation of jump tails. *Econometrica*, 79(6), pp.1727-1783.

Bollerslev, T. and Todorov, V., 2023. The jump leverage risk premium. *Journal of Financial Economics*, 150, 103772.

Bouchaud, J.P., Bonart, J., Donier, J. and Gould, M., 2018. *Trades, quotes and prices: financial markets under the microscope.* Cambridge University Press.

Brennan, M.J. and Subrahmanyam, A., 1996. Market microstructure and asset pricing: On the compensation for illiquidity in stock returns. *Journal of Financial Economics*, 41(3), pp.441-464.

Brunnermeier, M.K. and Pedersen, L.H., 2005. Predatory trading. *The Journal of Finance*, 60(4), pp.1825-1863.

Chan, W.H. and Maheu, J.M., 2002. Conditional jump dynamics in stock market returns. *Journal of Business & Economic Statistics*, 20(3), pp.377-389.

Chordia, T., Subrahmanyam, A. and Anshuman, V.R., 2001. Trading activity and expected stock returns. *Journal of Financial Economics*, 59(1), pp.3-32.

Chung, K.H. and Zhang, H., 2014. A simple approximation of intraday spreads using daily data. *Journal of Financial Markets*, 17, pp.94-120.

Constantinides, G.M., 1986. Capital market equilibrium with transaction costs. *Journal of Political Economy*, 94(4), pp.842-862.

Corsi, F., 2009. A simple approximate long-memory model of realized volatility. *Journal of Financial Econometrics*, 7, pp.174–196.

Cortez, K., Rodríguez-García, M. and Mongrutet, S., 2021. Exchange market liquidity prediction with the K-nearest neighbor approach: Crypto vs. fiat currencies. *Mathematics*, 9, 56.

Duan, J.C., Ritchken, P. and Sun, Z., 2006. Approximating GARCH-JUMP Models, Jump-Diffusion Processes, And Option Pricing. *Mathematical Finance*, 16(1), pp.21-52.

Dyhrberg, A.H., 2016. Bitcoin, gold and the dollar–A GARCH volatility analysis. *Finance Research Letters*, 16, pp.85-92.

Eraker, B., Johannes, M. and Polson, N., 2003. The impact of jumps in volatility and returns. *The Journal of Finance*, 58(3), pp.1269-1300.

Fong, K.Y., Holden, C.W. and Trzcinka, C.A., 2017. What are the best liquidity proxies for global research? *Review of Finance*, 21(4), pp.1355-1401.

Huang, M., 2003. Liquidity shocks and equilibrium liquidity premia. *Journal of Economic Theory*, 109(1), pp.104-129.

Hussain, S.M., 2011. The intraday behaviour of bid-ask spreads, trading volume and return volatility: evidence from DAX30. *International Journal of Economics and Finance*, 3(1), pp.23-34.

Liu, Y., Liu, H. and Zhang, L., 2019. Modeling and forecasting return jumps using realized variation measures. *Economic Modelling*, 76, pp.63-80.

Makarov, I. and Schoar, A., 2020. Trading and arbitrage in cryptocurrency markets. *Journal of Financial Economics*, 135(2), pp.293-319.

Manahov, V., 2021. Cryptocurrency liquidity during extreme price movements: is there a problem with virtual money? *Quantitative Finance*, 21, pp.341-360.

Pástor, Ľ. and Stambaugh, R.F., 2003. Liquidity risk and expected stock returns. *Journal of Political Economy*, 111(3), pp.642-685.

Pereira, J.P. and Zhang, H.H., 2010. Stock returns and the volatility of liquidity. *Journal of Financial and Quantitative Analysis*, 45(4), pp.1077-1110.

Shen, D., Urquhart, A., and Wang, P., 2022. Bitcoin intraday time series momentum. *The Financial Review*, 57, pp.319-344.

Vayanos, D., 1998. Transaction costs and asset prices: A dynamic equilibrium model. *The Review of Financial Studies*, 11(1), pp.1-58.

Vayanos, D., 2004. Flight to quality, flight to liquidity, and the pricing of risk. *NBER Working Paper* 10327.

Zhu, M., Song, Y. and Zheng, X., 2024. Volatility Dynamics and Mixed Jump-GARCH Model Based Jump Detection in Financial Markets. *Computational Economics*, online June 15, 2024.

# Table 1 – US stocks: Descriptive Statistics of Regular Return $r_t$, Liquidity-adjusted Return $r_t^\ell$, Liquidity Jump ($\beta_{r_t}^\ell$)

This table reports the daily descriptive statistics of all 15 US stocks over the entire sample period: Panel A –regular return ($r_t$); Panel B – liquidity-adjusted return ($r_t^\ell$); Panel C – liquidity jump ($\beta_{r_t}^\ell$). The value of $\beta_{r_t}^\ell$ is capped at 10.

| Panel A | daily regular return ($r_t$) | | | | | | | | | | | | | | |
|---|---|---|---|---|---|---|---|---|---|---|---|---|---|---|---|
| ticker | AAPL | AMZN | ATI | CMA | CRS | EME | GOOG | IBKR | LII | MLI | MSFT | NVDA | TPL | VFC | WSO |
| count | 2652 | 2652 | 2652 | 2652 | 2652 | 2652 | 2652 | 2652 | 2652 | 2652 | 2652 | 2652 | 2652 | 2652 | 2652 |
| mean | 0.10% | 0.12% | 0.07% | 0.04% | 0.09% | 0.11% | 0.08% | 0.11% | 0.08% | 0.09% | 0.10% | 0.23% | 0.16% | -0.01% | 0.07% |
| std | 1.78% | 2.05% | 3.39% | 2.54% | 2.99% | 1.90% | 1.78% | 1.96% | 1.69% | 2.17% | 1.70% | 3.51% | 2.93% | 2.44% | 1.63% |
| min | -13.16% | -14.40% | -19.90% | -27.67% | -24.12% | -19.19% | -11.05% | -10.50% | -9.12% | -23.08% | -14.75% | -89.87% | -25.91% | -14.07% | -10.42% |
| median | 0.10% | 0.11% | -0.03% | 0.06% | 0.06% | 0.11% | 0.11% | 0.11% | 0.11% | 0.03% | 0.08% | 0.24% | 0.13% | 0.07% | 0.10% |
| max | 12.12% | 14.15% | 31.21% | 19.93% | 18.06% | 14.42% | 15.97% | 15.54% | 11.55% | 20.74% | 14.48% | 29.81% | 26.58% | 26.99% | 16.67% |
| skewness | 0.00 | 0.34 | 0.60 | -0.61 | 0.00 | -0.60 | 0.23 | 0.09 | 0.07 | 0.25 | 0.11 | -5.99 | 0.36 | 0.58 | 0.24 |
| kurtosis | 5.36 | 6.23 | 9.31 | 13.88 | 6.16 | 14.08 | 6.60 | 3.69 | 4.03 | 14.58 | 8.12 | 167.51 | 10.37 | 11.38 | 8.39 |

| Panel B | daily liquidity-adjusted return ($r_t^\ell$) | | | | | | | | | | | | | | |
|---|---|---|---|---|---|---|---|---|---|---|---|---|---|---|---|
| ticker | AAPL | AMZN | ATI | CMA | CRS | EME | GOOG | IBKR | LII | MLI | MSFT | NVDA | TPL | VFC | WSO |
| count | 2652 | 2652 | 2652 | 2652 | 2652 | 2652 | 2652 | 2652 | 2652 | 2652 | 2652 | 2652 | 2652 | 2652 | 2652 |
| mean | 0.08% | 0.08% | 0.33% | -0.06% | 0.05% | 0.14% | 0.25% | 0.04% | 0.30% | 0.05% | 0.13% | 0.17% | 0.44% | -0.21% | 0.06% |
| std | 2.55% | 3.24% | 10.63% | 9.10% | 8.35% | 5.07% | 2.69% | 4.00% | 5.37% | 6.57% | 3.03% | 5.50% | 12.55% | 6.81% | 6.29% |
| min | -30.78% | -33.41% | -99.04% | -98.99% | -99.16% | -99.09% | -24.84% | -36.68% | -40.01% | -98.94% | -99.01% | -99.02% | -58.74% | -99.63% | -79.11% |
| median | 0.04% | 0.04% | 0.04% | 0.00% | 0.09% | 0.13% | 0.20% | 0.03% | 0.26% | 0.01% | 0.14% | 0.02% | -0.14% | 0.15% | 0.20% |
| max | 18.78% | 50.07% | 198.19% | 174.59% | 140.75% | 35.70% | 38.02% | 32.45% | 92.00% | 103.90% | 25.27% | 73.80% | 93.18% | 133.65% | 163.48% |
| skewness | -0.15 | 2.41 | 4.24 | 2.68 | -0.35 | -3.59 | 1.60 | 0.05 | 2.94 | -0.31 | -12.87 | -2.71 | 0.49 | -2.49 | 4.30 |
| kurtosis | 17.87 | 47.81 | 87.84 | 103.46 | 62.35 | 66.33 | 40.13 | 10.66 | 52.60 | 60.51 | 436.21 | 106.95 | 2.80 | 133.19 | 197.57 |

| Panel C | liquidity jump ($\beta_{r_t}^\ell$) | | | | | | | | | | | | | | |
|---|---|---|---|---|---|---|---|---|---|---|---|---|---|---|---|
| ticker | AAPL | AMZN | ATI | CMA | CRS | EME | GOOG | IBKR | LII | MLI | MSFT | NVDA | TPL | VFC | WSO |
| count | 2652 | 2652 | 2652 | 2652 | 2652 | 2652 | 2652 | 2652 | 2652 | 2652 | 2652 | 2652 | 2652 | 2652 | 2652 |
| mean | 1.20 | 1.32 | 1.09 | 1.09 | 0.99 | 0.83 | 1.39 | 1.09 | 0.82 | 0.90 | 1.26 | 1.25 | 0.44 | 1.06 | 0.71 |
| std | 1.55 | 1.69 | 1.74 | 1.77 | 1.74 | 1.54 | 1.88 | 1.79 | 1.54 | 1.72 | 1.65 | 1.64 | 1.06 | 1.61 | 1.36 |
| min | 0.00 | 0.00 | 0.00 | 0.00 | 0.00 | 0.00 | 0.00 | 0.00 | 0.00 | 0.00 | 0.00 | 0.00 | 0.00 | 0.00 | 0.00 |
| media | 0.83 | 0.87 | 0.57 | 0.58 | 0.47 | 0.39 | 0.85 | 0.55 | 0.38 | 0.39 | 0.83 | 0.82 | 0.20 | 0.61 | 0.33 |
| max | 10.00 | 10.00 | 10.00 | 10.00 | 10.00 | 10.00 | 10.00 | 10.00 | 10.00 | 10.00 | 10.00 | 10.00 | 10.00 | 10.00 | 10.00 |
| number of days $\beta_{r_t}^\ell >= 1$ | 804 | 1015 | 652 | 648 | 560 | 433 | 997 | 662 | 480 | 481 | 894 | 888 | 187 | 666 | 381 |
| as % of total number of days | 30.32% | 38.27% | 24.59% | 24.43% | 21.12% | 16.33% | 37.59% | 24.96% | 18.10% | 18.14% | 33.71% | 33.48% | 7.05% | 25.11% | 14.37% |

Note: The "number of days $\beta_{r_t}^\ell >= 1$" row gives the number of days with high liquidity jump ($\beta_{r_t}^\ell \geq 1$), and the "as % of total number of days" row gives the number of days with high liquidity jump ($\beta_{r_t}^\ell \geq 1$) as a percentage of the total number of trading days (2,652 days), for each of the stocks.

**Table 2 – US Stocks: Descriptive Statistics of Regular Volatility $\sigma_t$, Liquidity-adjusted Volatility $\sigma_t^\ell$, Liquidity Diffusion ($\beta_{\sigma_t}^\ell$)**

This table reports the daily descriptive statistics of all 15 US stocks over the entire sample period: Panel A – regular volatility ($\sigma_t$); Panel B – liquidity-adjusted volatility ($\sigma_t^\ell$); Panel C – liquidity jump ($\beta_{\sigma_t}^\ell$). The value of $\beta_{\sigma_t}^\ell$ is capped at 10.

| Panel A | | | | | | | daily regular volatility ($\sigma_t$) | | | | | | | | |
|---|---|---|---|---|---|---|---|---|---|---|---|---|---|---|---|
| ticker | AAPL | AMZN | ATI | CMA | CRS | EME | GOOG | IBKR | LII | MLI | MSFT | NVDA | TPL | VFC | WSO |
| count | 2652 | 2652 | 2652 | 2652 | 2652 | 2652 | 2652 | 2652 | 2652 | 2652 | 2652 | 2652 | 2652 | 2652 | 2652 |
| mean | 1.50% | 1.76% | 3.05% | 2.19% | 2.95% | 1.88% | 1.56% | 2.00% | 1.91% | 2.12% | 1.43% | 2.55% | 3.34% | 1.96% | 1.90% |
| std | 0.98% | 1.23% | 1.73% | 1.68% | 1.41% | 0.99% | 1.02% | 1.07% | 0.98% | 1.16% | 0.94% | 2.31% | 1.79% | 1.32% | 0.93% |
| min | 0.43% | 0.40% | 0.91% | 0.53% | 0.76% | 0.61% | 0.43% | 0.73% | 0.69% | 0.72% | 0.40% | 0.62% | 0.50% | 0.58% | 0.67% |
| median | 1.24% | 1.46% | 2.65% | 1.81% | 2.65% | 1.65% | 1.30% | 1.72% | 1.68% | 1.87% | 1.19% | 2.13% | 2.96% | 1.62% | 1.71% |
| max | 13.81% | 19.98% | 18.33% | 38.54% | 17.62% | 12.20% | 12.08% | 15.53% | 9.88% | 18.85% | 13.78% | 89.97% | 21.25% | 18.18% | 11.37% |

| Panel B | | | | | | | daily liquidity-adjusted volatility ($\sigma_t^\ell$) | | | | | | | | |
|---|---|---|---|---|---|---|---|---|---|---|---|---|---|---|---|
| ticker | AAPL | AMZN | ATI | CMA | CRS | EME | GOOG | IBKR | LII | MLI | MSFT | NVDA | TPL | VFC | WSO |
| count | 2652 | 2652 | 2652 | 2652 | 2652 | 2652 | 2652 | 2652 | 2652 | 2652 | 2652 | 2652 | 2652 | 2652 | 2652 |
| mean | 2.07% | 2.38% | 5.87% | 4.48% | 6.44% | 4.40% | 2.13% | 3.65% | 4.80% | 5.41% | 1.99% | 3.61% | 16.06% | 3.59% | 5.20% |
| std | 1.75% | 2.59% | 8.73% | 8.48% | 5.96% | 3.85% | 2.01% | 2.69% | 3.93% | 4.99% | 2.49% | 4.31% | 8.31% | 5.74% | 4.84% |
| min | 0.45% | 0.51% | 1.29% | 0.74% | 1.57% | 1.41% | 0.53% | 1.05% | 1.22% | 1.51% | 0.52% | 0.62% | 0.72% | 0.79% | 1.15% |
| median | 1.62% | 1.83% | 4.20% | 2.95% | 5.36% | 3.65% | 1.69% | 2.92% | 3.99% | 4.31% | 1.55% | 2.74% | 14.09% | 2.66% | 4.29% |
| max | 31.25% | 66.33% | 193.90% | 171.65% | 130.78% | 99.20% | 35.74% | 38.17% | 83.51% | 99.11% | 99.20% | 99.12% | 72.93% | 131.62% | 145.31% |

| Panel C | | | | | | | liquidity diffusion ($\beta_{\sigma_t}^\ell$) | | | | | | | | |
|---|---|---|---|---|---|---|---|---|---|---|---|---|---|---|---|
| ticker | AAPL | AMZN | ATI | CMA | CRS | EME | GOOG | IBKR | LII | MLI | MSFT | NVDA | TPL | VFC | WSO |
| count | 2652 | 2652 | 2652 | 2652 | 2652 | 2652 | 2652 | 2652 | 2652 | 2652 | 2652 | 2652 | 2652 | 2652 | 2652 |
| mean | 0.76 | 0.79 | 0.63 | 0.61 | 0.50 | 0.46 | 0.77 | 0.60 | 0.45 | 0.44 | 0.76 | 0.76 | 0.22 | 0.63 | 0.41 |
| std | 0.08 | 0.08 | 0.15 | 0.13 | 0.12 | 0.11 | 0.07 | 0.14 | 0.12 | 0.12 | 0.07 | 0.10 | 0.08 | 0.12 | 0.11 |
| min | 0.39 | 0.30 | 0.07 | 0.07 | 0.10 | 0.11 | 0.31 | 0.11 | 0.08 | 0.08 | 0.14 | 0.14 | 0.07 | 0.08 | 0.08 |
| median | 0.77 | 0.80 | 0.65 | 0.63 | 0.51 | 0.47 | 0.77 | 0.61 | 0.45 | 0.44 | 0.76 | 0.77 | 0.21 | 0.65 | 0.41 |
| max | 1.79 | 1.21 | 1.07 | 1.11 | 0.91 | 0.86 | 1.08 | 1.01 | 1.04 | 0.83 | 1.59 | 1.92 | 0.98 | 1.01 | 0.74 |
| number of days $\beta_{\sigma_t}^\ell >= 1$ | 14 | 15 | 1 | 2 | 0 | 0 | 6 | 1 | 1 | 0 | 10 | 29 | 0 | 1 | 0 |
| as % of total number of days | 0.53% | 0.57% | 0.04% | 0.08% | 0.00% | 0.00% | 0.23% | 0.04% | 0.04% | 0.00% | 0.38% | 1.09% | 0.00% | 0.04% | 0.00% |

Note: The "number of days $\beta_{\sigma_t}^\ell >= 1$" row gives the number of days with high liquidity diffusion ($\beta_{\sigma_t}^\ell \geq 1$), and the "as % of total number of days" row gives the number of days with high liquidity diffusion ($\beta_{\sigma_t}^\ell \geq 1$) as a percentage of the total number of trading days (2,652 days), for each of the stocks.

**Table 3 – Cryptos: Descriptive Statistics of Regular Return $r_t$, Liquidity-adjusted Return $r_t^\ell$, Liquidity Jump ($\beta_{r_t}^\ell$)**

This table reports the daily descriptive statistics of all 10 cryptocurrencies over the entire sample period: Panel A – regular return ($r_t$); Panel B – liquidity-adjusted return ($r_t^\ell$); Panel C – liquidity jump ($\beta_{r_t}^\ell$). The value of $\beta_{r_t}^\ell$ is capped at 10.

| Panel A | daily regular return ($r_t$) | | | | | | | | | |
|---|---|---|---|---|---|---|---|---|---|---|
| ticker | ADA | AVAX | BCH | BTC | ETC | ETH | LTC | SOL | UNI | XRP |
| count | 1577 | 1577 | 1577 | 1577 | 1577 | 1577 | 1577 | 1577 | 1577 | 1577 |
| mean | 0.45% | 0.58% | 0.27% | 0.23% | 0.31% | 0.28% | 0.17% | 0.56% | 0.31% | 0.27% |
| std | 9.64% | 11.15% | 6.45% | 3.54% | 7.25% | 5.45% | 4.97% | 6.99% | 6.45% | 6.26% |
| min | -26.82% | -37.49% | -36.24% | -15.38% | -47.54% | -27.74% | -36.75% | -42.25% | -33.86% | -47.22% |
| median | 0.00% | -0.11% | 0.07% | 0.02% | -0.05% | 0.12% | 0.15% | 0.02% | 0.05% | 0.08% |
| max | 324.00% | 317.57% | 136.60% | 44.98% | 164.46% | 135.76% | 53.09% | 107.83% | 55.89% | 73.10% |
| skewness | 24.11 | 16.88 | 7.03 | 1.73 | 7.88 | 9.76 | 0.70 | 2.63 | 1.77 | 1.81 |
| kurtosis | 805.06 | 445.50 | 135.18 | 21.20 | 171.80 | 243.27 | 12.71 | 38.24 | 13.14 | 26.64 |

| Panel B | daily liquidity-adjusted return ($r_t^\ell$) | | | | | | | | | |
|---|---|---|---|---|---|---|---|---|---|---|
| ticker | ADA | AVAX | BCH | BTC | ETC | ETH | LTC | SOL | UNI | XRP |
| count | 1577 | 1577 | 1577 | 1577 | 1577 | 1577 | 1577 | 1577 | 1577 | 1577 |
| mean | 2.07% | 2.64% | 1.31% | 0.86% | 0.81% | 1.30% | 1.98% | 1.99% | 1.97% | 2.02% |
| std | 18.41% | 14.20% | 8.75% | 4.58% | 8.85% | 10.03% | 7.31% | 6.95% | 6.38% | 7.40% |
| min | -20.71% | -42.11% | -55.41% | -26.39% | -75.12% | -49.82% | -41.30% | -36.71% | -45.60% | -67.73% |
| median | 0.98% | 1.84% | 0.85% | 0.41% | 0.35% | 0.69% | 1.63% | 1.24% | 1.31% | 1.22% |
| max | 699.25% | 495.22% | 276.08% | 107.20% | 255.26% | 355.89% | 221.07% | 128.23% | 73.67% | 71.47% |
| skewness | 34.55 | 26.96 | 19.83 | 8.77 | 15.11 | 28.06 | 17.11 | 4.62 | 1.81 | 1.41 |
| kurtosis | 1307.64 | 922.78 | 619.74 | 192.35 | 439.15 | 991.20 | 514.16 | 73.37 | 18.19 | 26.11 |

| Panel C | liquidity jump ($\beta_{r_t}^\ell$) | | | | | | | | | |
|---|---|---|---|---|---|---|---|---|---|---|
| ticker | ADA | AVAX | BCH | BTC | ETC | ETH | LTC | SOL | UNI | XRP |
| count | 1577 | 1577 | 1577 | 1577 | 1577 | 1577 | 1577 | 1577 | 1577 | 1577 |
| mean | 1.88 | 1.98 | 1.99 | 1.54 | 1.88 | 1.63 | 1.82 | 1.93 | 1.99 | 1.57 |
| std | 2.39 | 2.51 | 2.53 | 2.03 | 2.41 | 2.02 | 2.45 | 2.37 | 2.49 | 2.12 |
| min | 0.00 | 0.00 | 0.00 | 0.00 | 0.00 | 0.01 | 0.00 | 0.00 | 0.00 | 0.00 |
| median | 1.00 | 1.05 | 1.04 | 0.92 | 0.99 | 1.01 | 0.90 | 1.09 | 1.06 | 0.85 |
| max | 10 | 10 | 10 | 10 | 10 | 10 | 10 | 10 | 10 | 10 |
| number of days $\beta_{r_t}^\ell >= 1$ | 789 | 813 | 817 | 708 | 783 | 800 | 724 | 868 | 825 | 652 |
| as % of total number of days | 50.03% | 51.55% | 51.81% | 44.90% | 49.65% | 50.73% | 45.91% | 55.04% | 52.31% | 41.34% |

Note: The "number of days $\beta_{r_t}^\ell >= 1$" row gives the number of days with high liquidity jump ($\beta_{r_t}^\ell \geq 1$), and the "as % of total number of days" row gives the number of days with high liquidity jump ($\beta_{r_t}^\ell \geq 1$) as a percentage of the total number of trading days (1,577 days), for each of the cryptocurrencies.

**Table 4 – Cryptos: Descriptive Statistics of Regular Volatility $\sigma_t$, Liquidity-adjusted Volatility $\sigma_t^\ell$, Liquidity Diffusion ($\beta_{\sigma_t}^\ell$)**

This table reports the daily descriptive statistics of all 10 cryptocurrencies over the entire sample period: Panel A – regular volatility ($\sigma_t$); Panel B – liquidity-adjusted volatility ($\sigma_t^\ell$); Panel C – liquidity jump ($\beta_{\sigma_t}^\ell$). The value of $\beta_{\sigma_t}^\ell$ is capped at 10.

| Panel A | | | | daily regular volatility ($\sigma_t$) | | | | | | |
|---|---|---|---|---|---|---|---|---|---|---|
| ticker | ADA | AVAX | BCH | BTC | ETC | ETH | LTC | SOL | UNI | XRP |
| count | 1577 | 1577 | 1577 | 1577 | 1577 | 1577 | 1577 | 1577 | 1577 | 1577 |
| mean | 5.03% | 6.15% | 4.81% | 3.06% | 5.21% | 3.75% | 4.70% | 6.06% | 5.69% | 4.93% |
| std | 3.13% | 3.48% | 2.76% | 1.78% | 3.26% | 2.32% | 2.67% | 3.58% | 2.89% | 3.79% |
| min | 1.31% | 1.20% | 1.38% | 0.25% | 1.54% | 0.30% | 0.89% | 1.60% | 1.08% | 0.96% |
| median | 4.20% | 5.19% | 4.04% | 2.70% | 4.37% | 3.27% | 4.14% | 5.19% | 5.22% | 3.77% |
| max | 47.66% | 39.04% | 43.66% | 21.41% | 42.57% | 33.26% | 32.35% | 39.63% | 32.40% | 43.70% |

| Panel B | | | | daily liquidity-adjusted volatility ($\sigma_t^\ell$) | | | | | | |
|---|---|---|---|---|---|---|---|---|---|---|
| ticker | ADA | AVAX | BCH | BTC | ETC | ETH | LTC | SOL | UNI | XRP |
| count | 1577 | 1577 | 1577 | 1577 | 1577 | 1577 | 1577 | 1577 | 1577 | 1577 |
| mean | 5.58% | 6.22% | 4.54% | 3.71% | 5.00% | 4.39% | 5.19% | 6.47% | 5.43% | 5.76% |
| std | 4.24% | 4.27% | 3.53% | 2.23% | 4.04% | 2.98% | 3.21% | 4.34% | 3.36% | 4.61% |
| min | 0.63% | 0.95% | 0.78% | 0.57% | 0.98% | 0.58% | 0.98% | 0.80% | 0.85% | 0.93% |
| median | 4.59% | 4.98% | 3.64% | 3.21% | 3.80% | 3.77% | 4.45% | 5.46% | 4.89% | 4.40% |
| max | 86.85% | 58.21% | 75.46% | 30.68% | 77.51% | 54.81% | 49.63% | 59.55% | 43.71% | 59.90% |

| Panel C | | | | liquidity diffusion ($\beta_{\sigma_t}^\ell$) | | | | | | |
|---|---|---|---|---|---|---|---|---|---|---|
| ticker | ADA | AVAX | BCH | BTC | ETC | ETH | LTC | SOL | UNI | XRP |
| count | 1577 | 1577 | 1577 | 1577 | 1577 | 1577 | 1577 | 1577 | 1577 | 1577 |
| mean | 0.97 | 1.10 | 1.15 | 0.82 | 1.12 | 0.86 | 0.92 | 1.01 | 1.15 | 0.87 |
| std | 0.27 | 0.52 | 0.32 | 0.07 | 0.26 | 0.07 | 0.13 | 0.46 | 0.36 | 0.12 |
| min | 0.55 | 0.32 | 0.58 | 0.43 | 0.55 | 0.52 | 0.65 | 0.65 | 0.67 | 0.62 |
| median | 0.91 | 0.94 | 1.06 | 0.82 | 1.05 | 0.85 | 0.90 | 0.88 | 1.05 | 0.85 |
| max | 6.17 | 6.01 | 5.13 | 1.15 | 2.92 | 1.26 | 3.49 | 8.42 | 6.01 | 3.86 |
| number of days $\beta_{\sigma_t}^\ell >= 1$ | 413 | 607 | 968 | 15 | 940 | 77 | 252 | 358 | 924 | 124 |
| as % of total number of days | 26.19% | 38.49% | 61.38% | 0.95% | 59.61% | 4.88% | 15.98% | 22.70% | 58.59% | 7.86% |

Note: The "number of days $\beta_{\sigma_t}^\ell >= 1$" row gives the number of days with high liquidity diffusion ($\beta_{\sigma_t}^\ell \geq 1$), and the "as % of total number of days" row gives the number of days with high liquidity diffusion ($\beta_{\sigma_t}^\ell \geq 1$) as a percentage of the total number of trading days (1,577 days), for each of the cryptocurrencies.

## Table 5 – T-tests for Regular ARMA-GARCH Model and Liquidity-adjusted ARMA-GARCH Model for US Stocks
## Panel A – comparisons of model fit

Panel A of Table 5 reports the t-test results comparing model fit with the log-likelihood measure between the regular ARMA-GARCH model and liquidity-adjusted ARMA-GARCH for all selected US stocks.

| Panel A: log-likelihood regular vs. liquidity-adjusted | T | dof | alternative | p-val | alternative | p-val | alternative | p-val | sig | interpretation | direction |
|---|---|---|---|---|---|---|---|---|---|---|---|
| AAPL | 27.36 | 4818 | two-sided | 0.00 | less | 1.00 | greater | 0.00 | *** | liquidity adjustment does not improve model fit | ↓ |
| AMZN | 23.49 | 4818 | two-sided | 0.00 | less | 1.00 | greater | 0.00 | *** | liquidity adjustment does not improve model fit | ↓ |
| ATI | 69.03 | 4818 | two-sided | 0.00 | less | 1.00 | greater | 0.00 | *** | liquidity adjustment does not improve model fit | ↓ |
| CMA | 64.39 | 4818 | two-sided | 0.00 | less | 1.00 | greater | 0.00 | *** | liquidity adjustment does not improve model fit | ↓ |
| CRS | 95.73 | 4818 | two-sided | 0.00 | less | 1.00 | greater | 0.00 | *** | liquidity adjustment does not improve model fit | ↓ |
| EME | 131.33 | 4818 | two-sided | 0.00 | less | 1.00 | greater | 0.00 | *** | liquidity adjustment does not improve model fit | ↓ |
| GOOG | 23.45 | 4818 | two-sided | 0.00 | less | 1.00 | greater | 0.00 | *** | liquidity adjustment does not improve model fit | ↓ |
| IBKR | 83.37 | 4818 | two-sided | 0.00 | less | 1.00 | greater | 0.00 | *** | liquidity adjustment does not improve model fit | ↓ |
| LII | 99.63 | 4818 | two-sided | 0.00 | less | 1.00 | greater | 0.00 | *** | liquidity adjustment does not improve model fit | ↓ |
| MLI | 114.33 | 4818 | two-sided | 0.00 | less | 1.00 | greater | 0.00 | *** | liquidity adjustment does not improve model fit | ↓ |
| MSFT | 20.68 | 4818 | two-sided | 0.00 | less | 1.00 | greater | 0.00 | *** | liquidity adjustment does not improve model fit | ↓ |
| NVDA | 30.40 | 4818 | two-sided | 0.00 | less | 1.00 | greater | 0.00 | *** | liquidity adjustment does not improve model fit | ↓ |
| TPL | 232.40 | 4818 | two-sided | 0.00 | less | 1.00 | greater | 0.00 | *** | liquidity adjustment does not improve model fit | ↓ |
| VFC | 51.62 | 4818 | two-sided | 0.00 | less | 1.00 | greater | 0.00 | *** | liquidity adjustment does not improve model fit | ↓ |
| WSO | 96.20 | 4818 | two-sided | 0.00 | less | 1.00 | greater | 0.00 | *** | liquidity adjustment does not improve model fit | ↓ |

*** - significant at 1% level, ** - significant at 5% level, * significant at 10% level.

Notes:

1. The higher the value of log-likelihood, the better model fit.

2. If the p-value of "two-sided" alternative indicates the means are statistically not equal, we then evaluate the p-value of either "less " or "greater" to identify the impact of liquidity adjustment on model fit.

3. If the t-test result with "less" alternative is significant for a stock, it means the liquidity-adjusted model fits better than the regular model, and therefore liquidity adjustment improves model fit for the stock.

4. If the t-test result with "greater" alternative is significant for a stock, it means the liquidity-adjusted model does not fit better than the regular model, and therefore liquidity adjustment does not improve model fit for the stock.

5. The "direction" column indicates the effect of liquidity adjustment: ↑ indicates that liquidity adjustment improves the model fit, ↓ means liquidity adjustment does not improve model fit, and ↔ indicates liquidity adjustment has no statistically significant impact on model fit.

## Table 5 – T-tests for Regular ARMA-GARCH Model and Liquidity-adjusted ARMA-GARCH Model for US Stocks
### Panel B – comparisons between shock coefficients $a$ and $a^\ell$ in GARCH models

Panel B of Table 5 reports the t-test results comparing shock coefficient $a$ in the regular ARMA-GARCH model and shock coefficient $a^\ell$ in the liquidity-adjusted ARMA-GARCH for all selected US stocks. The values of $a$ and $a^\ell$ are shock coefficients in GARCH(1,1) specification in Equations 8 and 11.

$$regular: \sigma_t^{\omega 2} = \omega + a\epsilon_{t-1}^2 + b\sigma_{t-1}^{\omega\,2} \qquad (8)$$
$$liquidity\text{-}adjusted: \sigma_t^{\omega^{\ell 2}} = \omega^\ell + a^\ell \epsilon_{t-1}^{\ell\,2} + b^\ell \sigma_{t-1}^{\omega^{\ell 2}} \qquad (11)$$

| Panel B: $a$ vs. $a^\ell$ regular vs. liquidity-adjusted | T | dof | alternative | p-val | alternative | p-val | alternative | p-val | sig | interpretation | direction |
|---|---|---|---|---|---|---|---|---|---|---|---|
| AAPL | 10.70 | 4818 | two-sided | 0.00 | less | 1.00 | greater | 0.00 | *** | liquidity adjustment reduces shock coefficient | ↓ |
| AMZN | -1.19 | 4818 | two-sided | 0.23 | less | 0.12 | greater | 0.88 |  | liquidity adjustment has no significant impact on shock coefficient | ↔ |
| ATI | 6.26 | 4818 | two-sided | 0.00 | less | 1.00 | greater | 0.00 | *** | liquidity adjustment reduces shock coefficient | ↓ |
| CMA | 15.35 | 4818 | two-sided | 0.00 | less | 1.00 | greater | 0.00 | *** | liquidity adjustment reduces shock coefficient | ↓ |
| CRS | -2.29 | 4818 | two-sided | 0.02 | less | 0.01 | greater | 0.99 | ** | liquidity adjustment increases shock coefficient | ↑ |
| EME | 9.00 | 4818 | two-sided | 0.00 | less | 1.00 | greater | 0.00 | *** | liquidity adjustment reduces shock coefficient | ↓ |
| GOOG | 7.04 | 4818 | two-sided | 0.00 | less | 1.00 | greater | 0.00 | *** | liquidity adjustment reduces shock coefficient | ↓ |
| IBKR | 12.23 | 4818 | two-sided | 0.00 | less | 1.00 | greater | 0.00 | *** | liquidity adjustment reduces shock coefficient | ↓ |
| LII | 7.51 | 4818 | two-sided | 0.00 | less | 1.00 | greater | 0.00 | *** | liquidity adjustment reduces shock coefficient | ↓ |
| MLI | 5.78 | 4818 | two-sided | 0.00 | less | 1.00 | greater | 0.00 | *** | liquidity adjustment reduces shock coefficient | ↓ |
| MSFT | -2.04 | 4818 | two-sided | 0.04 | less | 0.02 | greater | 0.98 | ** | liquidity adjustment increases shock coefficient | ↑ |
| NVDA | -0.99 | 4818 | two-sided | 0.32 | less | 0.16 | greater | 0.84 |  | liquidity adjustment has no significant impact on shock coefficient | ↔ |
| TPL | 30.45 | 4818 | two-sided | 0.00 | less | 1.00 | greater | 0.00 | *** | liquidity adjustment reduces shock coefficient | ↓ |
| VFC | 6.34 | 4818 | two-sided | 0.00 | less | 1.00 | greater | 0.00 | *** | liquidity adjustment reduces shock coefficient | ↓ |
| WSO | 16.66 | 4818 | two-sided | 0.00 | less | 1.00 | greater | 0.00 | *** | liquidity adjustment reduces shock coefficient | ↓ |

\*\*\* - significant at 1% level, \*\* - significant at 5% level, \* significant at 10% level.

Notes:

1. If the p-value of "two-sided" alternative indicates the means are statistically not equal, we then evaluate the p-value of either "less " or "greater" to identify the impact of liquidity adjustment on model fit.

2. If the t-test result with "less" alternative is significant, it means $a$ is less than $a^\ell$, and therefore liquidity adjustment increases the shock efficient.

3. If the t-test result with "greater" alternative is significant, it means $a$ is greater than $a^\ell$, and therefore liquidity adjustment reduces the stock coefficient.

4. The "direction" column indicates the effect of liquidity adjustment: ↑ indicates that liquidity adjustment increases the shock coefficient, ↓ means liquidity adjustment reduces the shock coefficient, and ↔ indicates liquidity adjustment has no statistically significant impact on the shock coefficient.

## Table 5 – T-tests for Regular ARMA-GARCH Model and Liquidity-adjusted ARMA-GARCH Model for US Stocks
## Panel C – comparisons between volatility coefficients $b$ and $b^\ell$ in GARCH models

Panel C of Table 5 reports the t-test results comparing volatility coefficient $b$ in the regular ARMA-GARCH model and volatility coefficient $b^\ell$ in the liquidity-adjusted ARMA-GARCH for all selected US stocks. The values of $b$ and $b^\ell$ are shock coefficients in GARCH(1,1) specification in Equations 8 and 11.

$$regular: \sigma_t^{\omega 2} = \omega + a\epsilon_{t-1}^2 + b\sigma_{t-1}^{\omega\,2} \qquad (8)$$
$$liquidity\text{-}adjusted: \sigma_t^{\omega^{\ell 2}} = \omega^\ell + a^\ell \epsilon_{t-1}^{\ell\,2} + b^\ell \sigma_{t-1}^{\omega^{\ell 2}} \qquad (11)$$

| Panel C: $b$ vs. $b^\ell$ regular vs. liquidity-adjusted | T | dof | alternative | p-val | alternative | p-val | alternative | p-val | sig | interpretation | direction |
|---|---|---|---|---|---|---|---|---|---|---|---|
| AAPL | -7.12 | 4818 | two-sided | 0.00 | less | 0.00 | greater | 1.00 | *** | liquidity adjustment increases volatility coefficient | ↑ |
| AMZN | 2.12 | 4818 | two-sided | 0.03 | less | 0.98 | greater | 0.02 | ** | liquidity adjustment reduces volatility coefficient | ↓ |
| ATI | 6.27 | 4818 | two-sided | 0.00 | less | 1.00 | greater | 0.00 | *** | liquidity adjustment reduces volatility coefficient | ↓ |
| CMA | -11.48 | 4818 | two-sided | 0.00 | less | 0.00 | greater | 1.00 | ** | liquidity adjustment increases volatility coefficient | ↑ |
| CRS | 3.34 | 4818 | two-sided | 0.00 | less | 1.00 | greater | 0.00 | *** | liquidity adjustment reduces volatility coefficient | ↓ |
| EME | -1.22 | 4818 | two-sided | 0.22 | less | 0.11 | greater | 0.89 | | liquidity adjustment has no significant impact on volatility coefficient | ↔ |
| GOOG | -1.72 | 4818 | two-sided | 0.09 | less | 0.04 | greater | 0.96 | ** | liquidity adjustment increases volatility coefficient | ↑ |
| IBKR | -5.96 | 4818 | two-sided | 0.00 | less | 0.00 | greater | 1.00 | *** | liquidity adjustment increases volatility coefficient | ↑ |
| LII | 1.57 | 4818 | two-sided | 0.12 | less | 0.94 | greater | 0.06 | * | liquidity adjustment reduces volatility coefficient | ↓ |
| MLI | 1.12 | 4818 | two-sided | 0.26 | less | 0.87 | greater | 0.13 | | liquidity adjustment has no significant impact on volatility coefficient | ↔ |
| MSFT | -2.84 | 4818 | two-sided | 0.00 | less | 0.00 | greater | 1.00 | *** | liquidity adjustment increases volatility coefficient | ↑ |
| NVDA | -1.35 | 4818 | two-sided | 0.18 | less | 0.09 | greater | 0.91 | * | liquidity adjustment increases volatility coefficient | ↑ |
| TPL | -28.96 | 4818 | two-sided | 0.00 | less | 0.00 | greater | 1.00 | *** | liquidity adjustment increases volatility coefficient | ↑ |
| VFC | 3.37 | 4818 | two-sided | 0.00 | less | 1.00 | greater | 0.00 | *** | liquidity adjustment reduces volatility coefficient | ↓ |
| WSO | -1.91 | 4818 | two-sided | 0.06 | less | 0.03 | greater | 0.97 | ** | liquidity adjustment increases volatility coefficient | ↑ |

\*\*\* - significant at 1% level, \*\* - significant at 5% level, \* significant at 10% level.

Notes:

1. If the p-value of "two-sided" alternative indicates the means are statistically not equal, we then evaluate the p-value of either "less " or "greater" to identify the impact of liquidity adjustment on model fit.

2. If the t-test result with "less" alternative is significant, it means $b$ is less than $b^\ell$, and therefore liquidity adjustment increases the volatility efficient.

3. If the t-test result with "greater" alternative is significant, it means $b$ is greater than $b^\ell$, and therefore liquidity adjustment reduces the volatility coefficient.

4. The "direction" column indicates the effect of liquidity adjustment: ↑ indicates that liquidity adjustment increases the volatility coefficient, ↓ means liquidity adjustment reduces the volatility coefficient, and ↔ indicates liquidity adjustment has no statistically significant impact on the volatility coefficient.

## Table 6 – T-tests for Regular ARMA-GARCH Model and Liquidity-adjusted ARMA-GARCH Model for Cryptocurrencies
### Panel A – comparisons of model fit

Panel A of Table 6 reports the t-test results on model fit with the log-likelihood measure between the regular ARMA-GARCH model and liquidity-adjusted ARMA-GARCH for all selected cryptocurrencies.

| Panel A: log-likelihood regular vs. liquidity-adjusted | T | dof | alternative | p-val | alternative | p-val | alternative | p-val | sig | interpretation | direction |
|---|---|---|---|---|---|---|---|---|---|---|---|
| ADA | -3.40 | 2422 | two-sided | 0.00 | less | 0.00 | greater | 1.00 | *** | liquidity adjustment improves model fit | ↑ |
| AVAX | -0.66 | 2422 | two-sided | 0.51 | less | 0.25 | greater | 0.75 | | liquidity adjustment has no significant impact on model fit | ↔ |
| BCH | -3.79 | 2422 | two-sided | 0.00 | less | 0.00 | greater | 1.00 | *** | liquidity adjustment improves model fit | ↑ |
| BTC | 3.84 | 2422 | two-sided | 0.00 | less | 1.00 | greater | 0.00 | *** | liquidity adjustment does not improve model fit | ↓ |
| ETC | -5.07 | 2422 | two-sided | 0.00 | less | 0.00 | greater | 1.00 | *** | liquidity adjustment improves model fit | ↑ |
| ETH | -4.55 | 2422 | two-sided | 0.00 | less | 0.00 | greater | 1.00 | *** | liquidity adjustment improves model fit | ↑ |
| LTC | -1.03 | 2422 | two-sided | 0.30 | less | 0.15 | greater | 0.85 | | liquidity adjustment has no significant impact on model fit | ↔ |
| SOL | -8.20 | 2422 | two-sided | 0.00 | less | 0.00 | greater | 1.00 | *** | liquidity adjustment improves model fit | ↑ |
| UNI | -7.54 | 2422 | two-sided | 0.00 | less | 0.00 | greater | 1.00 | *** | liquidity adjustment improves model fit | ↑ |
| XRP | -13.29 | 2422 | two-sided | 0.00 | less | 0.00 | greater | 1.00 | *** | liquidity adjustment improves model fit | ↑ |

*** - significant at 1% level, ** - significant at 5% level, * significant at 10% level.

Notes:

1. The higher the value of log-likelihood, the better model fit.

2. If the p-value of "two-sided" alternative indicates the means are statistically not equal, we then evaluate the p-value of either "less " or "greater" to identify the impact of liquidity adjustment on model fit.

3. If the t-test result with "less" alternative is significant for a cryptocurrency, it means the liquidity-adjusted model fits better than the regular model, and therefore liquidity adjustment improves model fit for the cryptocurrency.

4. If the t-test result with "greater" alternative is significant for a cryptocurrency, it means the liquidity-adjusted model does not fit better than the regular model, and therefore liquidity adjustment does not improve model fit for the cryptocurrency.

5. The "direction" column indicates the effect of liquidity adjustment: ↑ indicates that liquidity adjustment improves the model fit, ↓ means liquidity adjustment does not improve model fit, and ↔ indicates liquidity adjustment has no statistically significant impact on model fit.

**Table 6 – T-tests for Regular ARMA-GARCH Model and Liquidity-adjusted ARMA-GARCH Model for Cryptocurrencies**
**Panel B – comparisons between shock coefficients $a$ and $a^\ell$ in GARCH models**

Panel B of Table 6 reports the t-test results comparing shock coefficient $a$ in the regular ARMA-GARCH model and shock coefficient $a^\ell$ in the liquidity-adjusted ARMA-GARCH for all selected US stocks. The values of $a$ and $a^\ell$ are shock coefficients in GARCH(1,1) specification in Equations 8 and 11.

$$regular: \sigma_t^{\omega 2} = \omega + a\epsilon_{t-1}^2 + b\sigma_{t-1}^{\omega\,2} \qquad (8)$$
$$liquidity\text{-}adjusted: \sigma_t^{\omega^{\ell 2}} = \omega^\ell + a^\ell \epsilon_{t-1}^{\ell\,2} + b^\ell \sigma_{t-1}^{\omega\,\ell 2} \qquad (11)$$

| Panel B: $a$ vs. $a^\ell$ regular vs. liquidity-adjusted | T | dof | alternative | p-val | alternative | p-val | alternative | p-val | sig | interpretation | direction |
|---|---|---|---|---|---|---|---|---|---|---|---|
| ADA | -3.45 | 2422 | two-sided | 0.00 | less | 0.00 | greater | 1.00 | *** | liquidity adjustment increases shock coefficient | ↑ |
| AVAX | -13.34 | 2422 | two-sided | 0.00 | less | 0.00 | greater | 1.00 | *** | liquidity adjustment increases shock coefficient | ↑ |
| BCH | -7.30 | 2422 | two-sided | 0.00 | less | 0.00 | greater | 1.00 | *** | liquidity adjustment increases shock coefficient | ↑ |
| BTC | -4.88 | 2422 | two-sided | 0.00 | less | 0.00 | greater | 1.00 | *** | liquidity adjustment increases shock coefficient | ↑ |
| ETC | -1.51 | 2422 | two-sided | 0.13 | less | 0.07 | greater | 0.93 | | liquidity adjustment has no significant impact on shock coefficient | ↔ |
| ETH | -17.43 | 2422 | two-sided | 0.00 | less | 0.00 | greater | 1.00 | *** | liquidity adjustment increases shock coefficient | ↑ |
| LTC | -18.30 | 2422 | two-sided | 0.00 | less | 0.00 | greater | 1.00 | *** | liquidity adjustment increases shock coefficient | ↑ |
| SOL | -10.95 | 2422 | two-sided | 0.00 | less | 0.00 | greater | 1.00 | *** | liquidity adjustment increases shock coefficient | ↑ |
| UNI | -27.36 | 2422 | two-sided | 0.00 | less | 0.00 | greater | 1.00 | *** | liquidity adjustment increases shock coefficient | ↑ |
| XRP | -3.46 | 2422 | two-sided | 0.00 | less | 0.00 | greater | 1.00 | *** | liquidity adjustment increases shock coefficient | ↑ |

\*\*\* - significant at 1% level, \*\* - significant at 5% level, \* significant at 10% level.

Notes:

1. If the p-value of "two-sided" alternative indicates the means are statistically not equal, we then evaluate the p-value of either "less " or "greater" to identify the impact of liquidity adjustment on model fit.

2. If the t-test result with "less" alternative is significant, it means $a$ is less than $a^\ell$, and therefore liquidity adjustment increases the shock efficient.

3. If the t-test result with "greater" alternative is significant, it means $a$ is greater than $a^\ell$, and therefore liquidity adjustment reduces the stock coefficient.

4. The "direction" column indicates the effect of liquidity adjustment: ↑ indicates that liquidity adjustment increases the shock coefficient, ↓ means liquidity adjustment reduces the shock coefficient, and ↔ indicates liquidity adjustment has no statistically significant impact on the shock coefficient.

# Table 6 – T-tests for Regular ARMA-GARCH Model and Liquidity-adjusted ARMA-GARCH Model for Cryptocurrencies
## Panel C – comparisons between volatility coefficients $b$ and $b^{\ell}$ in GARCH models

Panel C of Table 6 reports the t-test results comparing volatility coefficient $b$ in the regular ARMA-GARCH model and volatility coefficient $b^{\ell}$ in the liquidity-adjusted ARMA-GARCH for all selected US stocks. The values of $b$ and $b^{\ell}$ are shock coefficients in GARCH(1,1) specification in Equations 8 and 11.

$$regular: \sigma_t^{\omega 2} = \omega + a\epsilon_{t-1}^2 + b\sigma_{t-1}^{\omega\,2} \quad (8)$$
$$liquidity\text{-}adjusted: \sigma_t^{\omega^{\ell 2}} = \omega^{\ell} + a^{\ell}\epsilon_{t-1}^{\ell\,2} + b^{\ell}\sigma_{t-1}^{\omega^{\ell 2}} \quad (11)$$

| Panel C: $b$ vs. $b^{\ell}$ regular vs. liquidity-adjusted | T | dof | alternative | p-val | alternative | p-val | alternative | p-val | sig | interpretation | |
|---|---|---|---|---|---|---|---|---|---|---|---|
| ADA | 1.40 | 2422 | two-sided | 0.16 | less | 0.92 | greater | 0.08 | | liquidity adjustment has no significant impact on volatility coefficient | ↔ |
| AVAX | 13.18 | 2422 | two-sided | 0.00 | less | 1.00 | greater | 0.00 | *** | liquidity adjustment decreases volatility coefficient | ↓ |
| BCH | -0.83 | 2422 | two-sided | 0.41 | less | 0.20 | greater | 0.80 | | liquidity adjustment has no significant impact on volatility coefficient | ↔ |
| BTC | 7.65 | 2422 | two-sided | 0.00 | less | 1.00 | greater | 0.00 | *** | liquidity adjustment decreases volatility coefficient | ↓ |
| ETC | 0.14 | 2422 | two-sided | 0.89 | less | 0.55 | greater | 0.45 | | liquidity adjustment has no significant impact on volatility coefficient | ↔ |
| ETH | 16.25 | 2422 | two-sided | 0.00 | less | 1.00 | greater | 0.00 | *** | liquidity adjustment decreases volatility coefficient | ↓ |
| LTC | 18.45 | 2422 | two-sided | 0.00 | less | 1.00 | greater | 0.00 | *** | liquidity adjustment decreases volatility coefficient | ↓ |
| SOL | 2.71 | 2422 | two-sided | 0.01 | less | 1.00 | greater | 0.00 | *** | liquidity adjustment decreases volatility coefficient | ↓ |
| UNI | 17.33 | 2422 | two-sided | 0.00 | less | 1.00 | greater | 0.00 | *** | liquidity adjustment decreases volatility coefficient | ↓ |
| XRP | 4.81 | 2422 | two-sided | 0.00 | less | 1.00 | greater | 0.00 | *** | liquidity adjustment decreases volatility coefficient | ↓ |

*** - significant at 1% level, ** - significant at 5% level, * significant at 10% level.

Notes:

1. If the p-value of "two-sided" alternative indicates the means are statistically not equal, we then evaluate the p-value of either "less " or "greater" to identify the impact of liquidity adjustment on model fit.

2. If the t-test result with "less" alternative is significant, it means $b$ is less than $b^{\ell}$, and therefore liquidity adjustment increases the volatility efficient.

3. If the t-test result with "greater" alternative is significant, it means $b$ is greater than $b^{\ell}$, and therefore liquidity adjustment reduces the volatility coefficient.

4. The "direction" column indicates the effect of liquidity adjustment: ↑ indicates that liquidity adjustment increases the volatility coefficient, ↓ means liquidity adjustment reduces the volatility coefficient, and ↔ indicates liquidity adjustment has no statistically significant impact on the volatility coefficient.

## Table 7 – Performance Comparisons of MV Portfolios

This table compares the portfolio performance in Sharpe Ratio of a traditional mean variance (TMV) portfolio (Portfolio 1) and that of a liquidity-adjusted mean variance (LAMV) portfolio (Portfolio) for 15 US stocks (Panel A) and 10 cryptocurrencies (Panel B). The return of the risk-free asset ($r_f$) is assumed to be 0%. Both TMV and LAMV are optimized with the mean variance (MV) construct given in Equation 13.

$$\max_{W_t} \left( \hat{\mu}_{t+1}^{arga} W_t^{Tr} - \frac{\lambda_t}{2} W_t^{Tr} \bar{\Sigma}_t W_t \right) \quad (13)$$

| Panel A: stocks | | | | | | | | | Portfolio Sharpe Ratio (annualized, $r_f$ = 0%) | | | | | | |
|---|---|---|---|---|---|---|---|---|---|---|---|---|---|---|---|
| portfolio description | Portfolio Number | AAPL | AMZN | ATI | CMA | CRS | EME | GOOG | IBKR | LII | MLI | MSFT | NVDA | TPL | VFC | WSO |
| TMV | 1 | 0.79 | 0.62 | 0.23 | -0.02 | 0.37 | 0.70 | 0.49 | 0.69 | 0.59 | 0.19 | 0.48 | 0.34 | 0.64 | 0.12 | 0.17 |
| LAMV | 2 | 0.45 | 0.95 | -0.09 | 0.43 | 0.37 | 0.61 | 0.58 | 0.71 | 0.52 | 0.53 | 0.57 | 1.53 | 0.47 | 0.32 | 0.21 |
| Direction | | ↓ | ↑ | ↓ | ↑ | ↔ | ↓ | ↑ | ↑ | ↓ | ↑ | ↑ | ↑ | ↓ | ↑ | ↑ |

| Panel B: crypto asets | | | | | | Portfolio Sharpe Ratio (annualized, $r_f$ = 0%) | | | | | |
|---|---|---|---|---|---|---|---|---|---|---|---|
| portfolio description | Portfolio Number | ADA | AVAX | BCH | BTC | ETC | ETH | LTC | SOL | UNI | XRP |
| TMV | 1 | -0.62 | -0.10 | -0.54 | 0.03 | -0.11 | -0.66 | -0.64 | 0.11 | -0.23 | -0.57 |
| LAMV | 2 | -0.20 | -0.02 | 0.41 | -0.18 | -0.06 | -0.30 | -0.14 | 0.35 | -0.20 | 0.67 |
| Direction | | ↑ | ↑ | ↑ | ↓ | ↑ | ↑ | ↑ | ↑ | ↑ | ↑ |

Notes:
1. TMV Portfolio 1 (ARMA-GARCH-enhanced TMV portfolio: $\hat{\mu}_{t+1}^{arga}$ is the return vector of ARMA-GARCH forecasted $r_t$ values for day $t+1$, $\hat{\mu}_{r_{t+1}}^{arga}$; $\bar{\Sigma}_t$ is the covariance matrix of $r_t$'s for the rolling window, or $\bar{\Sigma}_{r_t}$, in the MV construct of Equation 13.

2. LAMV Portfolio 2 (ARMA-GARCH-enhanced LAMV portfolio): $\hat{\mu}_{t+1}^{arga}$ is the return vector of ARMA-GARCH forecasted $r_t^{\ell}$ values for day $t+1$, $\hat{\mu}_{r_{t+1}^{\ell}}^{arga}$; $\bar{\Sigma}_t$ is the covariance matrix of $r_t^{\ell}$'s for the rolling window, or $\bar{\Sigma}_{r_t^{\ell}}$, in the MV construct of Equation 13.

# Figure 1 – Histograms of Daily Liquidity Jump $\beta^\ell_{r_t}$ – Stocks

This figure provides the distributions (histograms) of daily liquidity magnitude Beta ($\beta^\ell_{r_t}$) for all 15 stocks over the entire sample period. The maximum value of $\beta^\ell_{r_t}$ is capped at 10.

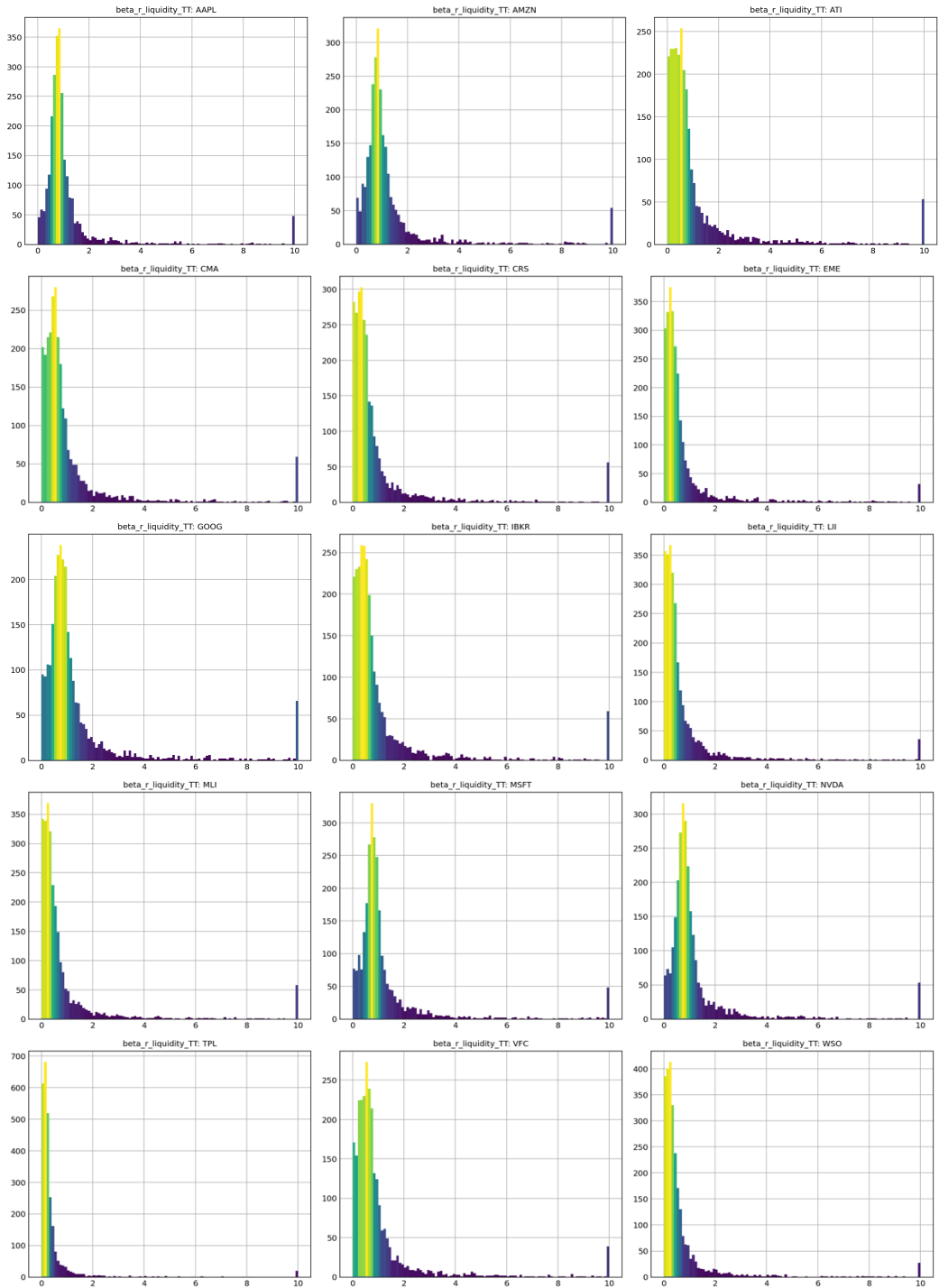

# Figure 2 – Histograms of Daily Liquidity Diffusion $\beta^{\ell}_{\sigma_t}$ – Stocks

This figure provides the distributions (histograms) of daily liquidity volatility Beta ($\beta^{\ell}_{\sigma_t}$) for all 15 stocks over the entire sample period.

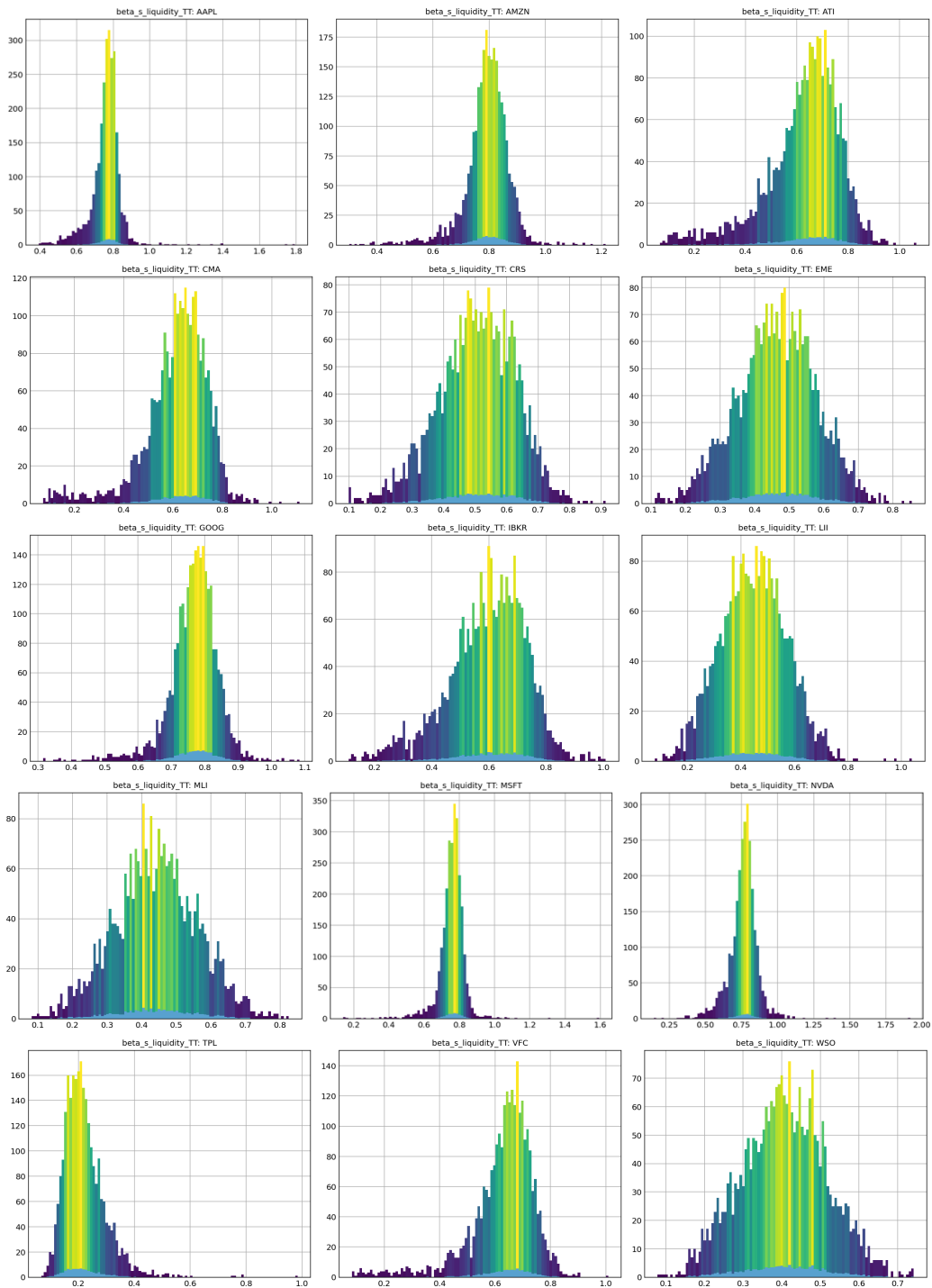

# Figure 3 – Histograms of Daily Liquidity Jump $\beta^{\ell}_{r_t}$ – Cryptocurrencies

This figure provides the distributions (histograms) of daily liquidity magnitude Beta ($\beta^{\ell}_{r_t}$) for all 10 cryptocurrencies over the entire sample period. The maximum value of $\beta^{\ell}_{r_t}$ is capped at 10.

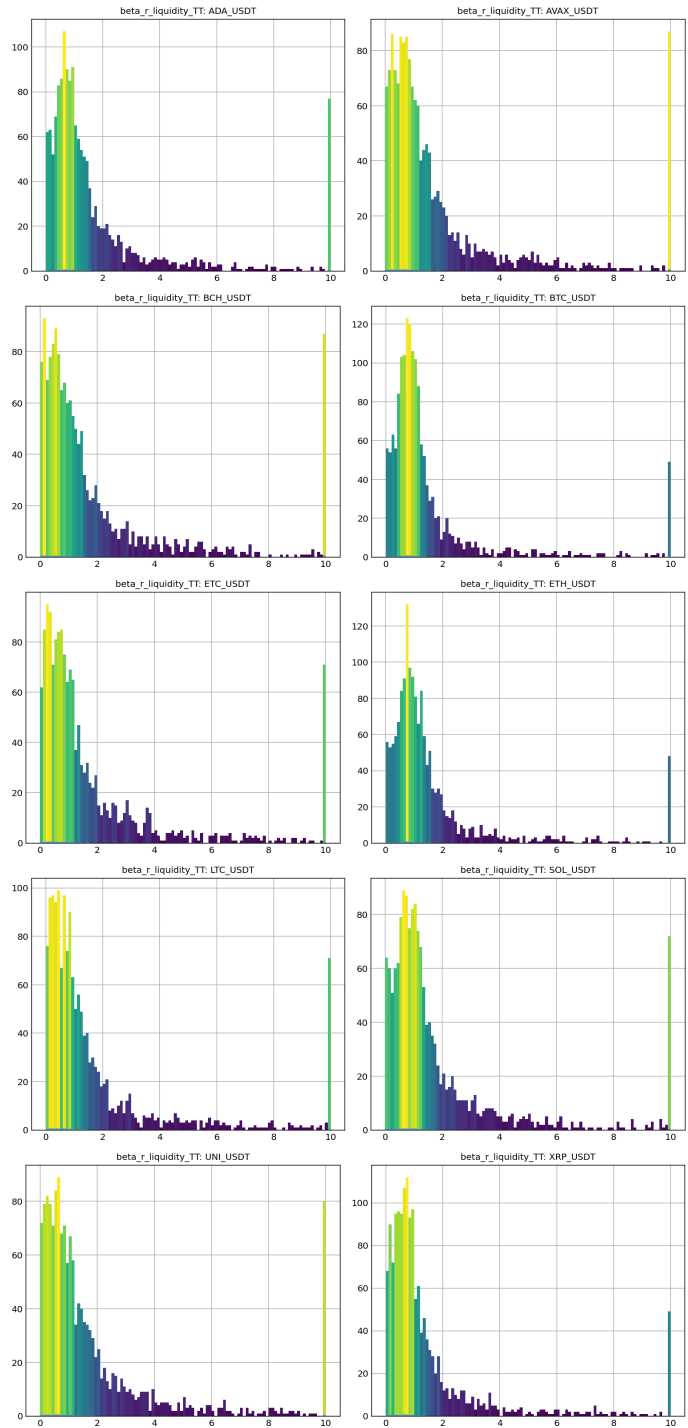

# Figure 4 – Histograms of Daily Liquidity Diffusion $\beta^{\ell}_{\sigma_t}$ – Cryptocurrencies

This figure provides the distributions (histograms) of daily liquidity volatility Beta ($\beta^{\ell}_{\sigma_t}$) for all 10 cryptocurrencies over the entire sample period.

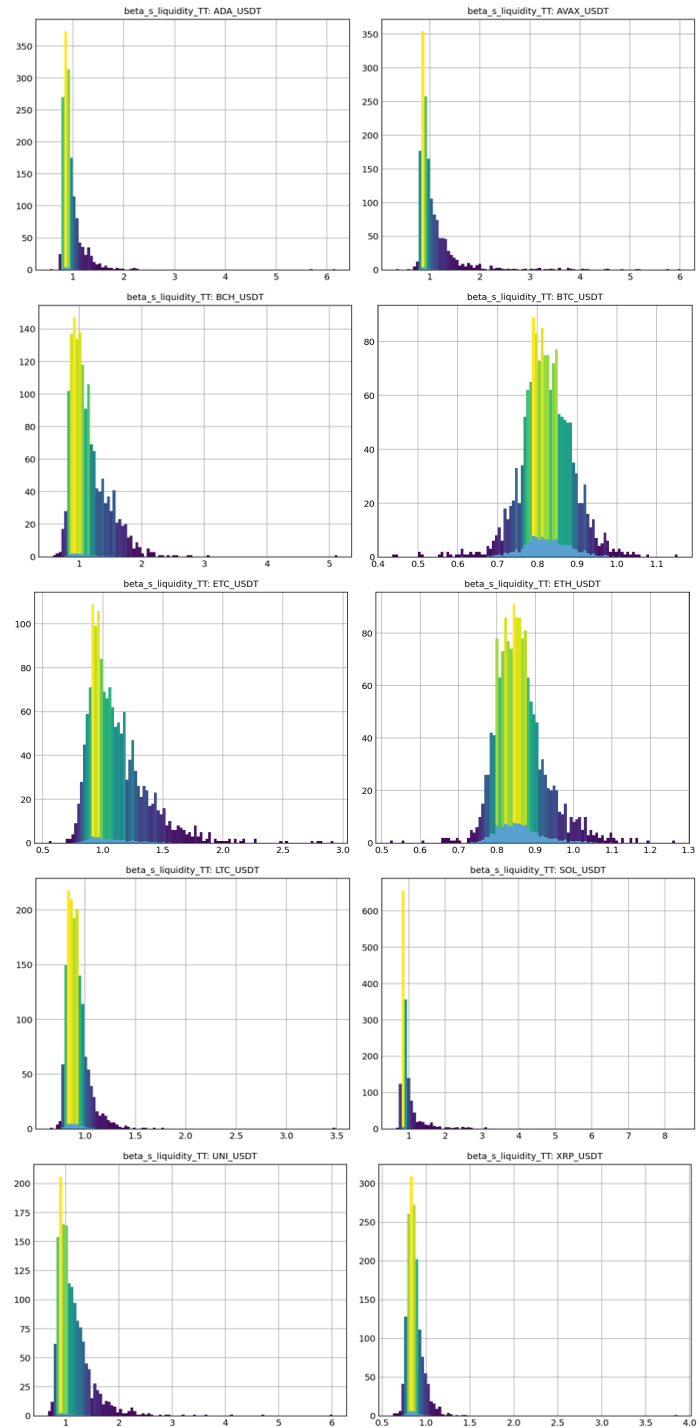